\pdfoutput=1
\documentclass[11pt,twoside,a4paper,cmspaper,final,collab]{cms-tdr}
\def\svnVersion{384e251}\def\svnDate{2025/08/22}\def\cmsCernNoTag{CERN-EP-2024-338}\def\cmsCernDate{\today}\def\cmsMessage{Published in the Journal of High Energy Physics as \href{http://dx.doi.org/10.1007/JHEP08(2025)085}{\doi{10.1007/JHEP08(2025)085}.}}
\begin{document}\cmsNoteHeader{EXO-22-014}

\newcommand{\ttjets}{\ensuremath{\ttbar}\xspace}
\newcommand{\tW}{\ensuremath{\PQt\PW}\xspace}
\newcommand{\tWDM}{\ensuremath{\PQt\PW}+DM\xspace}
\newcommand{\ttZ}{\ensuremath{\ttbar}\PZ\xspace}
\newcommand{\ttZtonunu}{\ensuremath{\ttbar \PZ \to \ttbar \PGn\PAGn}\xspace}
\newcommand{\tttwol}{\ensuremath{\ttbar(2\Pell)}\xspace}
\newcommand{\ttonel}{\ensuremath{\ttbar(1\Pell)}\xspace}
\newcommand{\Ztwol}{\ensuremath{\PZ(2\Pell)}\xspace}
\newcommand{\wjets}{\ensuremath{\PW}+jets\xspace}
\newcommand{\zjets}{\ensuremath{\PZ}+jets\xspace}
\newcommand{\Zinv}{\ensuremath{{\PZ} \to \PGn \PAGn}\xspace}
\newcommand{\ZLL}{\ensuremath{{\PZ} \to \Pell\PAell}\xspace}
\newcommand{\mchi}{\ensuremath{{m}_{\chi}}\xspace}
\newcommand{\tDM}{\ensuremath{{\PQt}/{\PAQt}}+DM\xspace}
\newcommand{\ttDM}{\ensuremath{\ttbar}+DM\xspace}
\newcommand{\njet}{\ensuremath{n_{\text{jet}}}\xspace}
\newcommand{\nlep}{\ensuremath{n_{\text{lep}}}\xspace}
\newcommand{\nbjet}{\ensuremath{n_{\PQb}}\xspace}
\newcommand{\HTratio}{\ensuremath{\pt(\mathrm{j_{1})}/\HT}\xspace}
\newcommand{\mindphi}{\ensuremath{\text{min }\Delta\phi(\mathrm{j_{1,2}}, \ptvecmiss)}\xspace}
\newcommand{\mTb}{\ensuremath{\mT^{\PQb}}\xspace}
\newcommand{\mTt}{\ensuremath{\mTii^{\PW}}\xspace}
\newcommand{\mTiill}{\ensuremath{\mTii^{\Pell\Pell}}\xspace}
\newcommand{\mll}{\ensuremath{m_{\Pell\Pell}}\xspace}

\newlength\cmsTabSkip\setlength{\cmsTabSkip}{2ex}
\providecommand{\cmsTable}[1]{\resizebox{\textwidth}{!}{#1}}

\cmsNoteHeader{EXO-22-014}

\title{Search for dark matter produced in association with one or two top quarks in proton-proton collisions at \texorpdfstring{$\sqrt{s} = 13\TeV$}{sqrt(s) = 13 TeV}}

\date{\today}

\abstract{
A search is performed for dark matter (DM) produced in association with a single top quark or a pair of top quarks using the data collected with the CMS detector at the LHC from proton-proton collisions at a center-of-mass energy of 13\TeV, corresponding to 138\fbinv of integrated luminosity. An excess of events with a large imbalance of transverse momentum is searched for across 0, 1 and 2 lepton final states. Novel multivariate techniques are used to take advantage of the differences in kinematic properties between the two DM production mechanisms. No significant deviations with respect to the standard model predictions are observed. The results are interpreted considering a simplified model in which the mediator is either a scalar or pseudoscalar particle and couples to top quarks and to DM fermions. Axion-like particles that are coupled to top quarks and DM fermions are also considered. Expected exclusion limits of 410 and 380\GeV for scalar and pseudoscalar mediator masses, respectively, are set at the 95\% confidence level.  A DM particle mass of 1\GeV is assumed, with mediator couplings to fermions and DM particles set to unity. A small signal-like excess is observed in data, with the largest local significance observed to be 1.9 standard deviations for the 150\GeV pseudoscalar mediator hypothesis. Because of this excess, mediator masses are only excluded below 310 (320)\GeV for the scalar (pseudoscalar) mediator. The results are also translated into model-independent 95\% confidence level upper limits on the visible cross section of DM production in association with top quarks, ranging from 1\unit{pb} to 0.02\unit{pb}.}

\hypersetup{
pdfauthor={CMS Collaboration},
pdftitle={Search for dark matter produced in association with one or two top quarks in proton-proton collisions at sqrt(s)=13 TeV},
pdfsubject={CMS},
pdfkeywords={CMS, dark matter, top quark}
}

\maketitle

\section{Introduction}
\label{sec:introduction}
Astrophysical observations point to the existence of dark matter (DM)~\cite{Porter:2011nv}. Its existence has been inferred from gravitational effects on galaxies and other large-scale objects. While the nature of DM remains elusive, there are many models suggesting that DM could be explained in the context of particle physics~\cite{Bertone:2004pz}. In a large class of these new theories, DM is made of stable weakly interacting massive particles that interact with the standard model (SM) constituents through mediator particles, and hence may be produced at colliders, such as the CERN Large Hadron Collider (LHC). The DM particles are effectively invisible to the detectors located at the LHC interaction points, traversing these experiments undetected and leaving behind an unbalanced momentum in the plane transverse to the proton beams. 

A convenient approach to model the coupling mechanisms between the SM and DM sectors is provided by simplified models, which can cover the main kinematic features exhibited in more complex ones. One possible simplification assumes the existence of either a new neutral scalar ($\phi$) or pseudoscalar (\Pa) particle that can interact with both the SM fermion sector and a new fermionic Dirac DM particle ($\chi$)~\cite{Abercrombie:2015wmb,PhysRevD.91.015017}. This construction involves a Yukawa-like coupling of the new mediator to fermions.

\begin{figure}[!htbp]
\centering
\includegraphics[width=0.32\textwidth]{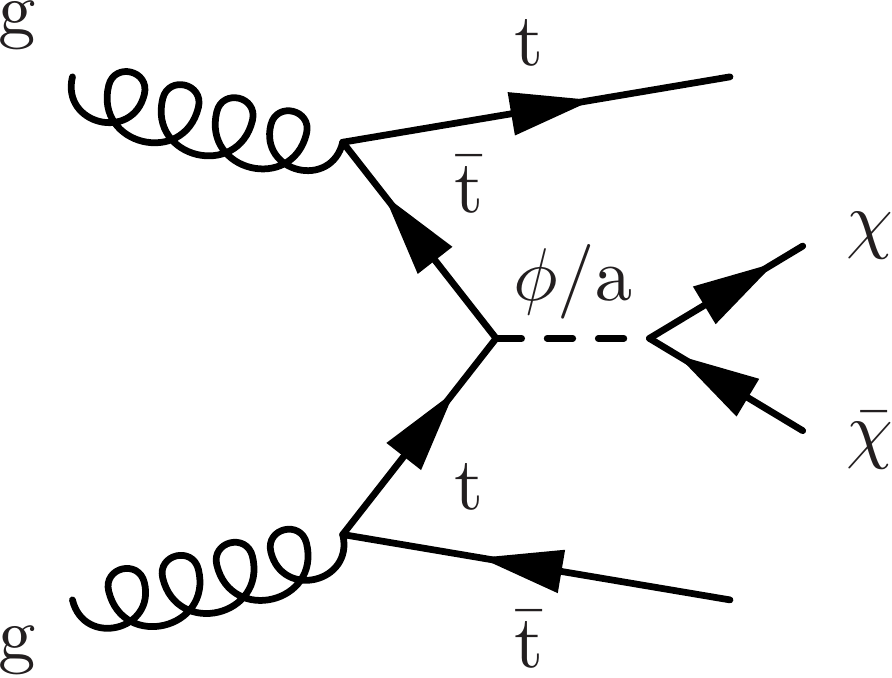}
\includegraphics[width=0.32\textwidth]{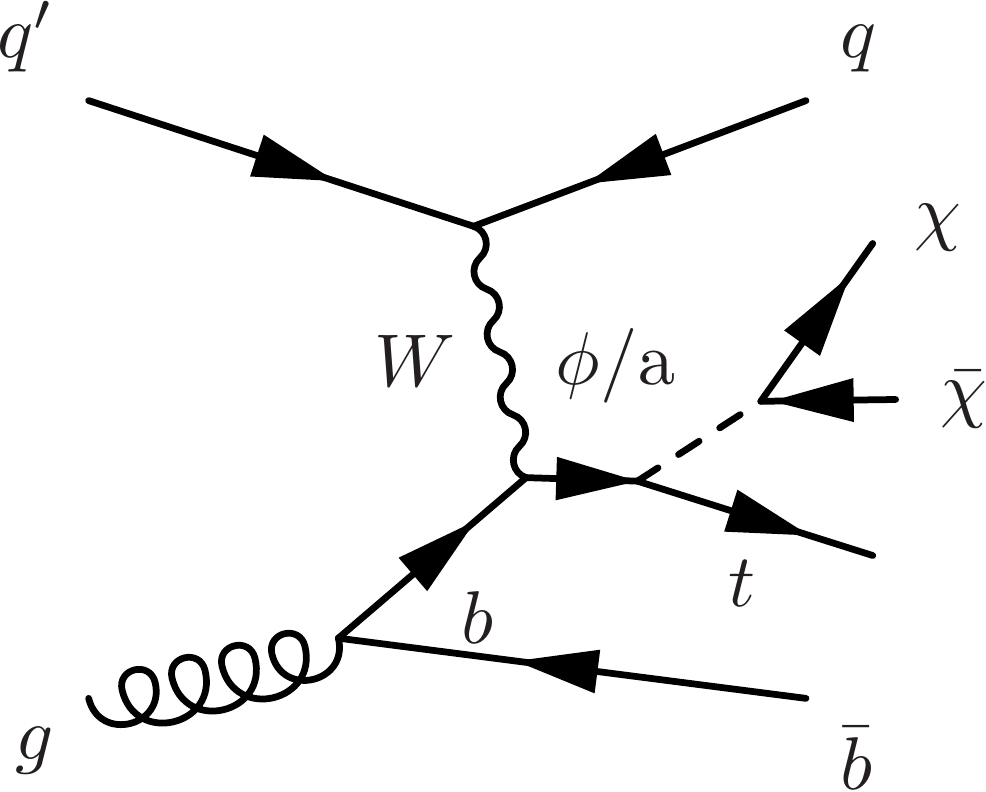}
\includegraphics[width=0.32\textwidth]{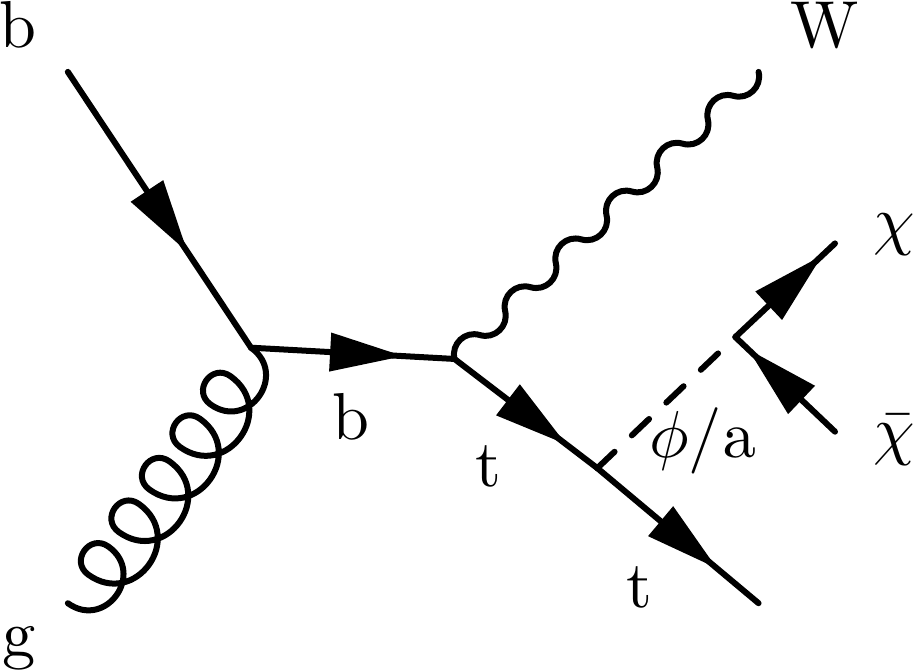}
\caption{Principal production diagrams in the context of the simplified model with a scalar/pseudoscalar ($\phi/\Pa$) mediator for the associated production of a pair of DM particles ($\chi$) with a top quark pair (left) and a single top quark in both $t$-channel (center), and tW-channel (right) production modes. The additional quark \PQq in the $t$-channel diagram is often produced at high pseudorapidity.}
\label{fig:intro:simp-model}
\end{figure}

The Lagrangians of these models can be expressed as follows~\cite{PhysRevD.91.015017}:
\begin{eqnarray}
\label{eq:lagr_scalar}
    \mathcal{L}_{\phi} = \mathcal{L}_{\textrm{SM}} 
    + \frac12(\partial_\mu \phi)^2 - \frac12 m_{\phi}^{2}\phi^{2} 
    + i\bar{\chi} {\partial\!\!\!\big /}\chi - m_{\chi} \bar{\chi}\chi
    -g_{\chi}\phi\bar{\chi}\chi
    -\frac{\phi}{\sqrt{2}}\sum_{q=\PQu,\PQd,\PQs,\PQc,\PQb,\PQt}g_{\PQq}y_{\PQq}q\bar{q},\\
\label{eq:lagr_pseudoscalar}
    \mathcal{L}_{\Pa} =  \mathcal{L}_{\textrm{SM}}
    + \frac12(\partial_\mu \Pa)^2 - \frac12 m_\Pa^{2}\Pa^{2} 
    + i\bar{\chi} {\partial\!\!\!\big /}\chi - m_{\chi} \bar{\chi}\chi
    -i g_{\chi}\Pa\bar{\chi}\gamma_{5}\chi -\frac{ia}{\sqrt{2}}\sum_{q=\PQu,\PQd,\PQs,\PQc,\PQb,\PQt}g_{\PQq}y_{\PQq}\bar{q}\gamma_{5}q,
\end{eqnarray}
where $\mathcal{L}_{\textrm{SM}}$ is the SM Lagrangian, and $y_{\PQq}$ is the SM Yukawa coupling, normalized to the Higgs vacuum expectation value as $y_{\PQq} = \sqrt{2}m_{\PQq}/\upsilon$. There are only four additional parameters that control the kinematic properties or cross section: the mass of the DM particle ($m_{\chi}$), the mass of the mediator ($m_{\phi/\Pa}$), the universal quark-mediator coupling ($g_{\textrm{\cPq}}$), and the DM-mediator coupling ($g_{\chi}$). Under these conditions, the mediators would couple preferentially to heavy third-generation quarks, which motivates the focus on DM particles produced in association with a top quark pair (\ttDM events) or a single top quark (\tDM events)~\cite{PhysRevD.96.035031}. The main processes involving the \ttDM and \tDM production in the context of this simplified model are shown in Fig.~\ref{fig:intro:simp-model}.

Complete models can also lead to enhanced couplings between the heavy quarks and the various DM mediators. One such case is the type II two-Higgs doublet model extended by an additional light pseudoscalar boson a (2HDM+\Pa)~\cite{LHCDarkMatterWorkingGroup:2018ufk,Bauer:2017ota}, featuring a differentiable coupling of scalar or pseudoscalar mass eigenstates to up-type and down-type fermions for a type II 2HDM configuration, and in which, for relatively low values of $\tan{\beta}$, the production of top quarks accompanied by DM particles is favored. This scenario is characterized by a higher number of parameters that control the kinematic properties with respect to the simplified model presented above and it is not considered in this paper. The simplified model considered in this analysis has a similar behavior to the 2HDM+\Pa model in the limit of high additional Higgs boson masses when the charged Higgs bosons are completely decoupled.

Further models that could provide a mediator to the dark sector are those that propose axions or axion-like particles (ALPs, denoted as $\textrm{A}$ in the following), which are spin-0 particles that do not interact with the SM gauge groups. The ALPs are a common feature in many extensions of the SM, where they emerge as pseudo-Nambu-Goldstone bosons because of an approximate axion shift symmetry. This symmetry allows for the masses of ALPs to be significantly smaller than the energy scale of the underlying ultraviolet model, making them promising candidates for studies at the LHC. Initially, axions were proposed as a solution to the strong-$CP$ problem~\cite{Peccei:1977hh, Peccei:1977ur, Weinberg:1977ma, Wilczek:1977pj}. The ALPs are also promising candidates for mediators to the dark sector~\cite{Dolan:2014ska,Bharucha:2022lty,Allen:2024ndv, Blasi:2023hvb}. These models often feature a Dirac fermionic DM candidate $\chi$ and a pseudoscalar mediator $\textrm{A}$ that interact with SM fermions. Since the couplings of ALPs to fermions are typically proportional to the fermion masses, the ALP coupling to top quarks becomes particularly significant. In this paper, we also exploit our analysis to search for ALPs as DM mediators using a Lagrangian of the form:
\begin{equation}
    \mathcal{L}_\textrm{A} = \mathcal{L}_{\textrm{SM}} + \frac{1}{2} (\partial_\mu \textrm{A}
    ) (\partial^\mu \textrm{A} ) -\frac12 m_\textrm{A}^2 \textrm{A}^2
    + i\bar{\chi} {\partial\!\!\!\big /}\chi - m_{\chi} \bar{\chi}\chi
    - i  c_\textrm{t} \, \frac{m_\textrm{t}}{f_\textrm{A}} \bar t  \gamma_5 \, t  \, \textrm{A}
    - i c_{\chi} \, \frac{m_{\chi}}{f_\textrm{A}} \bar{\chi}\gamma_{5}\chi \,\textrm{A}\, ,
    \label{eq:lagr_ALPs_pseudoscalar}
\end{equation}
where $f_\textrm{A}$ denotes the ALP decay constant, $c_{t}$ the ALP coupling to top quarks, and $c_{\chi}$ its coupling to DM particles. The similarity of this Lagrangian to the one in Eq.~(\ref{eq:lagr_pseudoscalar}) ensures that the DM signatures considered in Fig.~\ref{fig:intro:simp-model} can also be interpreted in terms of models with an ALP mediator $\textrm{A}$.

Searches for DM in signatures containing a top quark pair or a single top quark have been previously carried out by the ATLAS and CMS Collaborations at a center-of-mass energy of 13\TeV. The CMS Collaboration has performed a search that specifically targets the \ttDM production mode using the data set collected in 2016--2018 (Run 2) and covering all three commonly explored lepton categories (considering only muons and electrons)~\cite{CMS:2021eha}. These lepton categories are named in some previous works as $0\Pell$, $1\Pell$, and $2\Pell$ channels; here they will be referred to as all-hadronic (AH), single-lepton (SL), and dileptonic (DL) channels, respectively. The ATLAS Collaboration has delivered a combination across all channels using Run 2 data and considering both the \ttDM and \tDM production modes, optimizing for \ttDM topologies~\cite{ATLAS-CONF-2022-007}, with a further combination with an update to the SL channel, only considering the \ttDM signal~\cite{ATLAS:2024rcx}. Further, ATLAS analyses have searched for \tDM final states, in boosted topologies predicted by 2HDM+\Pa models~\cite{ATLAS:2022znu,ATLAS:2024kpy}. The work presented in this paper improves upon the constraints set by the CMS Collaboration with 2016 data from Run 2 on the combination of both the \ttDM and \tDM modes~\cite{CMS-PAS-EXO-18-010}. Various analysis improvements, as explained in later Sections, and the new dileptonic final state introduced lead to improved results of about 20\% with respect to the increase in sensitivity from solely adding the data available from 2017 and 2018. 

The analysis described in this paper is performed using the data collected with the CMS detector in 2016--2018, corresponding to an integrated luminosity of $138\fbinv$. Although it probes similar signal topologies to the searches described above, the analysis includes considerable differences in strategy for some of the channels. In this work, a search optimized for both \ttDM and \tDM processes mediated by the presence of a neutral spin-0 particle is performed across all channels. The analysis makes use of dedicated search strategies for both DM production mechanisms to further enhance the separation power against the SM background.
Novel multivariate techniques are employed using characteristic features of both signal production modes. An interpretation in the context of the spin-0 simplified DM model for both types of mediators is provided. 

The paper is organized as follows: a brief introduction of the CMS detector and event reconstruction is presented in Section~\ref{sec:cms-detector}, followed by a description in Section~\ref{sec:samples} of the data and of the simulated samples used. The event selection is presented in Section~\ref{sec:Selection}, while Section~\ref{sec:systematics} is dedicated to the treatment of experimental and theoretical systematic uncertainties in the analysis. The signal extraction methodology is discussed in Section~\ref{sec:fit}, and the results obtained after this procedure are shown in Section~\ref{sec:results}. The paper is summarized in Section~\ref{sec:summary}. 
Tabulated results are provided in the HEPData record for this analysis~\cite{hepdata}.

\section{The CMS detector and event reconstruction}
\label{sec:cms-detector}

The central feature of the CMS apparatus is a superconducting solenoid of 6\unit{m} internal diameter, providing a magnetic field of 3.8\unit{T}. Within the solenoid volume are a silicon pixel and strip tracker, a lead tungstate crystal electromagnetic calorimeter (ECAL), and a brass and scintillator hadron calorimeter (HCAL), each composed of a barrel and two endcap sections. Forward calorimeters extend the pseudorapidity ($\eta$) coverage provided by the barrel and endcap detectors. Muons are detected in gas-ionization chambers embedded in the steel flux-return yoke outside the solenoid. A more detailed description of the CMS detector, together with a definition of the coordinate system used and the relevant kinematic variables, can be found in Ref.~\cite{Chatrchyan:2008zzk}.

Events of interest are selected using a two-tiered trigger system~\cite{Khachatryan:2016bia}. The first level, composed of custom hardware processors, uses information from the calorimeters and muon detectors to select events at a rate of around 100\unit{kHz} within a time interval of less than 4\mus. The second level, known as the high-level trigger, consists of a farm of processors running a version of the full event reconstruction software optimized for fast processing, and reduces the event rate to around 1\unit{kHz} before data storage.

The particle-flow (PF) algorithm~\cite{CMS-PRF-14-001} aims to reconstruct and identify each individual particle in an event, with an optimized combination of information from the various elements of the CMS detector. The energy of photons is obtained directly from the ECAL measurement. The energy of electrons is obtained from a combination of the electron momentum at the primary interaction vertex as determined by the tracker, the energy of the corresponding ECAL cluster, and the energy sum of all bremsstrahlung photons spatially compatible with originating from the electron track. The muon track is obtained from the combination of central tracker and muon system information, and its curvature provides an estimate of the momentum. The energy of charged hadrons is determined from a combination of their momentum measured in the tracker and the matching ECAL and HCAL energy deposits, corrected for the response function of the calorimeters to hadronic showers. Finally, neutral hadrons are identified as HCAL energy clusters not linked to any charged-hadron trajectory, or as a combined ECAL and HCAL energy excess with respect to the expected charged-hadron energy deposit.

The primary vertex (PV) is taken to be the vertex corresponding to the hardest scattering in the event, evaluated using tracking information alone, as described in Section 9.4.1 of Ref.~\cite{CMS-TDR-15-02}.
For each event, hadronic jets are clustered from the particles reconstructed with PF (PF candidates) using the infrared- and collinear-safe anti-\kt algorithm~\cite{Cacciari:2008gp, Cacciari:2011ma} with a distance parameter of 0.4. The jet momentum is determined as the vector sum of all particle momenta in the jet and is found from simulation to be within 5--10\% of the parton's generated momentum over the entire \pt spectrum and detector acceptance. Additional proton-proton ($\Pp\Pp$) interactions within the same or nearby bunch crossings (pileup) can contribute with additional tracks and calorimetric energy depositions to the jet momentum. To mitigate this effect, tracks identified as originating from pileup vertices are discarded and an offset correction is applied to correct for remaining contributions~\cite{Cacciari:2008gn}. Jet energy corrections are derived from simulation and applied to calibrate the jet momentum. In situ measurements of the momentum balance in dijet, photon+jet, \zjets, and multijet events are used to account for any residual differences in jet energy scale in data and simulation~\cite{CMS-PAS-JME-16-003}. Additional selection criteria are applied to each jet to remove jets potentially dominated by anomalous contributions from various subdetector components or reconstruction failures~\cite{CMS-PAS-JME-16-003}.

The deep neural network (DNN)-based combined secondary vertex algorithm ({DeepCSV}) is used to identify jets originating from the hadronization of bottom quarks~\cite{CMS:2017wtu}, denoted in the following as ``{\PQb}-tagged jets''. At the operating point of the tagging algorithm chosen for this analysis, the efficiency of identifying {\PQb} quark jets in simulate  \ttbar events is about 75\%, averaged over all \pt, and the misidentification rate for light-flavor quark and gluon jets is about 1\%. Scale factors are applied to the simulated samples to reproduce the {\PQb} tagging efficiency measured in data.

The missing transverse momentum vector \ptvecmiss is defined as the negative vector \pt sum of all PF particles originating from the primary vertex; its magnitude is defined as \ptmiss. Jet energy scale and resolution corrections are also propagated to the \ptvecmiss calculation.

\section{Data and simulated samples}
\label{sec:samples}

The data used in this search were collected by the CMS detector in 2016--2018 and correspond to an integrated luminosity of 138\fbinv. Several triggers were used to collect the data, either requiring large \ptmiss or the presence of one or two leptons (electrons or muons).
The \ptmiss-based triggers are employed to select events that do not contain leptons (\ie, for the AH channel) if they have \ptmiss and missing hadronic activity \mht ~\cite{Khachatryan:2016bia} both above 120\GeV, where missing hadronic activity is defined as the negative vector sum of jet transverse momenta. These triggers are nearly 100\% efficient for events with \ptmiss of at least 250\GeV.
A second set of triggers, used in the SL channel, requires the presence of at least one isolated electron (muon) with $\pt>27$ (24)\GeV. The corresponding trigger efficiencies are above 90\% for leptons with $\pt>30\GeV$. The last set of triggers, for the DL channel, uses a combination of single- and di-lepton triggers to maximize the efficiency, which is on average 98\% for events with two leptons with $\pt > 25 (20)\GeV$.

Monte Carlo (MC) simulated samples of the main SM backgrounds and of the signal processes are used to optimize the event selection and to improve the background estimation. The SM processes with the largest contributions to the background in the various final states are \ttjets, \wjets, and \zjets production, although the precise composition depends on the specific channel under consideration. Simulated events of \ttjets production and single top quark processes ($t$-, $s$-, and \tW-channel) are generated at next-to-leading order (NLO) in quantum chromodynamics (QCD) using the {\POWHEG v2}~\cite{Nason:2004rx,Frixione:2007vw,Alioli:2010xd,Alioli:2009je,Re:2010bp,Frixione:2007nw} event generators. For the \ttjets process, the top quark \pt distribution is reweighted to match next-to-NLO (NNLO) theoretical computations~\cite{Czakon:2017wor}. The samples for top quark pair associated production with a \PW boson (\ensuremath{\ttbar}+{\PW}) and a \PZ boson (\ensuremath{\ttbar}+\PZ) plus up to two additional partons are generated based on the NLO ME calculations implemented in \MGvATNLO~and the FxFx~\cite{Frederix:2012ps} prescription to merge multileg processes. Samples of \zjets and \wjets are generated using leading order (LO) predictions with up to four partons in the final state using \MGvATNLO~\cite{Alwall:2014hca} (v2.2.2 in 2016, v2.4.2 in 2017 and 2018) with the MLM prescription~\cite{Mangano:2006rw} for matching jets from the matrix element (ME) calculation to the parton shower description. Dedicated electroweak~\cite{Denner:2009gj, Denner:2011vu, Denner:2012ts, Kuhn:2005gv, Kallweit:2014xda, Kallweit:2015dum} and QCD (calculated with \MGvATNLO) LO to NLO corrections ($K$ factors), parameterized as functions of the generated boson \pt, are applied to \zjets and \wjets events. Multijet events are simulated at LO either using \PYTHIA or \MGvATNLO event generators, making use of the MLM matching for the latter case. Other processes with a smaller contribution, such as diboson production, are generated at NLO using either \MGvATNLO using FxFx matching or with {\POWHEG v2} and normalized to the most accurate cross section calculations available~\cite{Gehrmann:2014fva,Campbell:1999ah}.

The \ttDM and \tDM signal processes are simulated at LO with \MGvATNLO v2.6.1 with one and zero additional partons, respectively, in the matrix element calculations using the simplified model referenced in Section~\ref{sec:introduction}.
The benchmark values chosen for the coupling parameters are $g_{\chi}=g_{\textrm{\cPq}}=1$, following the recommendations of the LHC Dark Matter Working Group~\cite{Boveia:2016mrp}. In addition, the mass of the DM candidate is set to $\mchi=1\GeV$ as a benchmark point, taking into account that for the on-shell production of the mediators, the impact of the DM mass in the \ptmiss distribution is irrelevant. For the off-shell production of $\phi/\Pa$, there is typically a moderate change in the distribution. However, these scenarios are not considered in this study because of their highly suppressed cross sections. The parameter scan focuses on variations of the parameter $m_{\phi/\Pa}$, chosen from the following list of mass points (for both type of mediators and both \ttDM and \tDM production modes): 50, 100, 150, 200, 250, 300, 350, 400, 450, and 500\GeV. 

The initial-state partons are modeled with the NNPDF 3.0~\cite{Ball:2014uwa} parton distribution function (PDF) set, of the same order in QCD as was used for the matrix element calculation for the samples corresponding to the 2016 period, while the NNPDF 3.1 NNLO~\cite{NNPDF:2017mvq} PDF set is used for the 2017 and 2018 samples. Parton showering and hadronization are handled by {\PYTHIA v8.226 (8.230)}~\cite{Sjostrand:2014zea} using the {CUETP8M1(2)} underlying event tune~\cite{Khachatryan:2015pea} for samples in the 2016 period, and the \textsc{CP5} tune~\cite{CMS:2019csb} for most samples corresponding to the 2017 and 2018 periods. All signal and background samples are processed using \GEANTfour~\cite{Agostinelli:2002hh} to provide a full simulation of the CMS detector, including a simulation of the previously mentioned triggers.
The effects of additional $\Pp\Pp$ interactions in the same or adjacent bunch crossings, referred to as pileup, are included in all simulation samples. To match the simulation distribution of genuine pileup interactions with the one observed in data, a reweighting procedure is implemented.
Correction factors are derived and applied to the simulated samples to match the trigger efficiencies measured in data. Additional corrections are applied to cover remaining residual differences between data and simulation that arise from the lepton identification and reconstruction efficiencies, as well as from {\PQb}-tagged jet identification efficiencies.

\section{Event selection}
\label{sec:Selection}
This search defines several orthogonal signal regions (SRs) that are statistically combined in a simultaneous global fit of the \ptmiss spectrum of the events in those regions. Control regions (CRs) enriched in the major background processes are included in the fit for each channel to improve the estimates of the SM contributions. Events are classified into three mutually exclusive ``channels'', based on the number of leptons in the final state: the AH channel, which contains events with no electrons or muons with $\pt>10\GeV$; the SL channel, which contains events with one lepton with $\pt>35 (30)\GeV$ for electrons (muons) and with no other lepton with $\pt>10\GeV$; and the DL channel, which contains events with one lepton with $\pt>25\GeV$, a second lepton with $\pt>20\GeV$, and no other lepton with $\pt>10\GeV$.

In the AH CRs and SL channel, electrons are selected if they have $\pt>35\GeV$ and $\abs{\eta}<2.1$, and muons are selected if they have $\pt>30\GeV$ and $\abs{\eta}<2.4$. For the DL channel, electrons and muons are selected if they have $\pt>20\GeV$ and $\abs{\eta}<2.4$. In all channels, events containing additional leptons with $\pt>10\GeV$ and $\abs{\eta}<2.4$ are vetoed. To ensure that candidate leptons are well measured, identification requirements based on hit information in the tracker and muon systems and on energy deposits in the calorimeters are imposed. Leptons are further required to be isolated from hadronic activity to reject leptons within jets that could arise, for example, from the decay of \PQb quarks.
A relative isolation quantity is defined as the scalar \pt sum of all PF candidates within a $\Delta{R}=\sqrt{\smash[b]{(\Delta\eta)^{2}+(\Delta\phi)^{2}}}$ cone of radius 0.3 (0.4) centered around the electron (muon) candidate, where $\phi$ is the azimuthal angle in radians, divided by the lepton. The requirement put on this relative isolation depends on the $\pt$ for electrons and is on average required to be less than $0.2$, while is required to be less than $0.15$ for muons~\cite{Khachatryan:2015hwa,CMS-PAPER-MUO-10-004}.

Each channel SR is further divided based on the number of {\PQb}-tagged jets (\nbjet) into a region with exactly one {\PQb}-tagged jet (1 \PQb), targeting the \tDM process, and a region with two or more {\PQb}-tagged jets (2 \PQb), targeting the \ttDM process. The regions with exactly one {\PQb}-tagged jet in the AH and SL channels are further subdivided into exactly zero or $\geq 1$ forward jet regions (0FJ and 1FJ) to increase the sensitivity to the $t$-channel \tDM process, which tends to have a forward jet, as shown in Fig.~\ref{fig:intro:simp-model}. Jet candidates are required to have $\pt > 30\GeV$ in the AH and SL channels, or $\pt > 20\GeV$ in the DL channel, and are categorized as ``central'' if they lie within $\abs{\eta} < 2.4$ and as ``forward'' if they are within $2.4< \abs{\eta} <4.0$. The \PQb-tagged jets are required to have the same $\pt$ threshold and to lie within $\abs{\eta} < 2.4$.

In each channel, a set of variables is identified to discriminate between signal and background events. In the AH and SL channels, a selection on these variables is applied to increase the signal significance, which is optimized using as a figure of merit the ratio between the number of expected signal events and the square root of the expected SM background contributions. In the DL channel, the discriminating variables are employed to train a neural network. 

\subsection{All-hadronic signal regions}
\label{sec:Selection_AH}

Events categorized into the AH channel are events with no electrons or muons with $\pt > 10\GeV$, at least one identified {\PQb}-tagged jet, at least three jets with $\pt > 30\GeV$, and $\ptmiss > 250\GeV$. The dominant background after this selection consists of SL \ttbar (\ttonel) events, where the lepton either falls out of detector acceptance or is not identified, providing an additional source of \ptmiss. 
To reduce this background, we require the transverse mass $\mTb$ of the \ptvecmiss vector and of a {\PQb}-tagged jet to be greater than 140\GeV. The $\mTb$ variable is defined as:
\begin{equation}
\mTb = \sqrt{\smash[b]{2\ptmiss \pt^{b}[1-\cos(\Delta\phi)]}},
\label{eq:mT_AH}
\end{equation}

where $\pt^{\PQb}$ is the transverse momentum of the {\PQb}-tagged jet and $\Delta\phi$ is the opening angle between the {\PQb}-tagged jet direction and the \ptvecmiss in the transverse plane. For \ttbar background events, \mTb tends to have values below or around the top quark mass in the case where the {\PQb}-tagged jet belongs to the top quark decay chain and the lepton is not identified. For the calculation, we choose the {\PQb}-tagged jet with the highest {DeepCSV} discriminant value if there is more than one candidate.

Studies have shown that to further reduce the \ttonel background, together with that from \Zinv events, an effective variable to use is \HTratio, which is defined as the ratio of the leading \pt jet in the event to \HT,  the scalar \pt sum of all reconstructed jets in the event with $\pt > 30\GeV$ within $\abs{\eta}<2.4$.
The \ttDM process, which has six jets at the ME level, tends to have lower values of this observable than the \ttonel and \Zinv backgrounds, which have fewer jets. We require $\HTratio < 0.5$ in the $\nbjet\geq 2$ category. The \tDM events, which have fewer central jets, tend to exhibit a distribution similar to that of the background, so no such requirement on the \HTratio is applied in the $\nbjet = 1$ category.

Another background in this channel is QCD multijet production. For this process, no intrinsic \ptmiss is expected, so the observed \ptmiss is mostly the result of jet mismeasurements.
For these events, \ptvecmiss is often aligned with one of the leading jets. As a result, the minimum opening angle between each of the two leading jets and the \ptvecmiss is required to satisfy $\mindphi>0.8$ radians to reduce this background.

The dominant backgrounds after the selection presented in Table~\ref{tab:SR} arise from \ttbar, \wjets, and \Zinv processes, with QCD multijet events, single top quark, Drell--Yan (DY), and diboson production giving smaller contributions.

\subsection{Single-lepton signal regions}
\label{sec:Selection_SL}
Events in the SL channel are selected by requiring the presence of one electron (muon) with $\pt>35 (30)\GeV$, ${\geq}1$ identified {\PQb}-tagged jets, at least two jets with $\pt>30\GeV$, and $\ptmiss>250\GeV$. After this selection, the dominant backgrounds are from \ttbar and \wjets processes.

To reduce the dominant backgrounds and improve the signal sensitivity, we impose a requirement on the transverse mass \mT, which is calculated as:
\begin{equation}
\mT = \sqrt{\smash[b]{2\ptmiss \pt^{\Pell}[1-\cos(\Delta\phi)]}},
\label{eq:mT_SL}
\end{equation}

where $\pt^{\Pell}$ is the transverse momentum of the lepton and $\Delta\phi$ is the opening angle between the lepton direction and the \ptvecmiss in the transverse plane. The \mT variable is expected to be less than the {\PW} boson mass for leptonic on-shell {\PW} decays in \ttbar and \wjets events, while for signal, off-shell {\PW} decays, and dileptonic decays of \ttbar, the \mT variable is expected to exceed the {\PW} boson mass because of the additional \ptmiss in the event. A requirement of $\mT>140\GeV$ reduces the background contributions from single-lepton events and enhances the analysis sensitivity to the DM signal.

After applying the \mT selection, the remaining \ttbar contributions are primarily from events with both top quarks decaying leptonically (\tttwol) and with one lepton not identified. This background can be further reduced by making use of the \mTt variable~\cite{Bai:2012gs}, which is defined as the minimal value of the mass of a particle assumed to be pair produced that decays to a {\PW} boson and a {\PQb} quark jet. The {\PW} bosons are assumed to be produced on-shell and to decay leptonically, where one of the two leptons is not detected. Based on the variable definition, in \tttwol events the \mTt distribution has a kinematic end point at the top quark mass for perfect detector response, while this is not the case for signal events, where two additional DM particles are present. The calculation of \mTt requires two {\PQb}-tagged jets, where one comes from the same decay chain as the reconstructed lepton. If only one {\PQb}-tagged jet is present in the event, each of the first three (or two in three-jet events) leading non-{\PQb}-tagged jets is assumed as the second {\PQb}-tagged jet in the calculation. The \mTt is then evaluated for all possible jet-lepton combinations and the minimum \mTt value is considered to discriminate between signal and background events.

If two or more {\PQb}-tagged jets are identified in the events, \mTt is calculated using all possible jet-lepton combinations. The smallest of all the \mTt values is taken as the event discriminant. A requirement of $\mTt > 180\GeV$ is applied in the analysis.

In addition, jets and the \ptvecmiss vector tend to be more separated in the transverse plane in signal events than in \ttbar background processes. To improve the search sensitivity, the minimum opening angle \mindphi in the transverse plane between the direction of each of the first two leading-\pt jets with $\abs{\eta}<2.4$ and the \ptvecmiss vector is required to be greater than $0.8$ radians.

To reduce the \ttbar background further, the transverse mass \mTb is required to be greater than 140\GeV.

To reject the \tttwol background, the modified topness variable $t$ \cite{Graesser:2012modifiedtopness} is introduced. Unlike \mTt, which minimizes the transverse mass of the invisible particles, the modified topness minimizes the reconstructed center-of-mass energy of the event, subject to the conditions that both top quarks and \PW bosons are on shell. It is defined as:
\begin{equation}
t = \ln(\text{min}\ S)
\end{equation} 
with 
\begin{equation}
\begin{aligned}
S = \frac{(m_{\PW}^2 - (p_{\PGn} + p_{\Pell})^2)^2}{a_{\PW}^4} + \frac{(m_{\PQt}^2 - (p_{\PQb}+p_{\PW})^2)^2}{a_{\PQt}^4},
\end{aligned}
\label{eq:had:modifiedtopness2}
\end{equation}
where $p$ and $m$ denote the four-momentum and mass of the particles involved, respectively. The minimization of the variable $S$ is done with respect to all components of the three momentum $\vec{p}_{\PW}$ and the component of the three momentum $\vec{p}_{\PGn}$ along the beam line. The sum is performed over all five assumed final-state particles from the top quark and \PW boson decays. The $a_{\PW} = 5\GeV$ and $a_{\PQt} = 15\GeV$ parameters determine the relative weighting of the mass shell conditions. The inputs to the calculation of $S$ are two jets, a lepton, and the \ptmiss. The value of $t$ quantifies how well an event is compatible with the \tttwol hypothesis. Rather than defining a single selection criterion on this variable, we split all the SRs in the SL channel into two further categories $t \leq 0$ (T1) and $t > 0$ (T2).

Though significantly reduced by these selections, \ttbar production, in particular \tttwol, remains the dominant background in this channel, followed by \wjets, single top quark, diboson, and DY production.

\subsection{Dileptonic signal regions}
\label{sec:Selection_DL}

Events in the dileptonic channel are selected by requiring at least two leptons with $\pt > 25$ and $20\GeV$, and at least one {\PQb}-tagged jet with $\pt > 30\GeV$.
The invariant mass \mll of the two-lepton system is required to satisfy $\mll > 20\GeV$ to suppress light resonances, and for the case of two electrons or two muons, the region $76 < \mll < 106\GeV$ is also excluded to suppress DY production. This selection gives a region that is dominated by \ttbar processes, which, because of the presence of two neutrinos in the final state, has a fairly hard \ptmiss spectrum that cannot be simply removed with a \ptmiss requirement. A selection is therefore imposed on \mTiill   \cite{Lester:1999, Khachatryan:2015vra}, which is defined as:
\begin{equation}
\mTiill=\underset{\vec{p}_{\text{T},\bar{\PGn}}+\vec{p}_{\text{T},\PGn}=\ptvecmiss}{\min}[\max\{M_{\text{T}}(m_{\Pell},m_{\bar{\PGn}},\vec{p}_{\text{T},\Pell},\vec{p}_{\text{T},\bar{\PGn}}),M_{\text{T}}(m_{\PAell},m_{\PGn},\vec{p}_{\text{T},\PAell},\vec{p}_{\text{T},\PGn})\}],
\label{eq:mt2ll:1}
\end{equation}
with 
\begin{equation}
M_{\text{T}}(m_{1},m_{2},\vec{p}_{\text{T},1},\vec{p}_{\text{T},2}) = \sqrt{m^2_{1}+m^2_{2}+2(E_{\text{T},1}E_{\text{T},2}-\vec{p}_{\text{T},1}\cdot\vec{p}_{\text{T},2})},
\label{eq:dil:mt2ll:2}
\end{equation}
where $m_{i}$, $\vec{p}_{\text{T},i}$, and $E_{\text{T},i}$ correspond to the mass, transverse momentum vector, and transverse energy of the particle $i$ respectively, while \ptvecmiss is the measured missing transverse momentum vector. This variable offers information on the mass of pair-produced particles in situations where both particles decay to a final state with undetected particles, as in the case of two leptonically decaying \PW bosons produced from top quark pairs. If the visible components in the decay chain are measured correctly, this variable has a kinematic endpoint at the \PW boson mass for the \ttbar processes but not for signal, because of the additional \ptmiss from DM. 

After requiring $\mTiill > 80\GeV$, the main remaining backgrounds are from the \ttbar production because of \ptmiss mismeasurements, \tW production, DY, and \ttZ events, where the \PZ boson decays to neutrinos (\ttZtonunu), giving a very similar signature to the signal process. Most of the backgrounds in the DL channel are well modeled by the MC simulation. However, the DY process is difficult to model as it normally only enters the selection if \ptmiss, and hence the \mTiill, are significantly mismodeled. The DL signal regions are therefore split by lepton flavor into same-flavor (SF) regions with two electrons or two muons, which will be relatively enriched in the DY process, and different flavor (DF) regions with one electron and one muon, which will be depleted in DY events. A method is then used to estimate the DY in the more enriched \tDM SF region (as discussed in Section~\ref{sec:DLCR}), based on control samples in data.

\subsubsection{Neural network optimization}
\label{sec:DL_NN}

As previously mentioned, for the DL channel the \ptmiss distribution alone does not provide a strong discrimination of the signal from the background. The signal sensitivity in the DL channel can be improved by employing a neural network (NN) trained on a number of discriminating variables. A separate NN is trained in each of the 1 \PQb tag and 2 \PQb tag regions (the SF and DF lepton regions are combined to maximize the number of events used for the training) and for each of the mediator hypotheses (scalar or pseudoscalar). Regarding the \tDM and \ttDM processes, both types of signals are considered in the 1 \PQb tag and 2 \PQb tag SRs since a large fraction of \ttDM events can contribute to the 1 \PQb tag region if one of the \PQb jets is not tagged. Only one NN is trained for all mediator mass hypotheses since the performance of the NN is found not to significantly improve when training on a single mass point compared to the ensemble of all mass points.

Various distributions are considered as inputs for the NN, and the following variables are found to improve the signal sensitivity (in addition to \ptmiss and \mTiill, mentioned previously):

\begin{itemize}
\item $\abs{\Delta\phi(\Pell,\PAell)}$, the opening angle between the two leptons. Since top quarks decay to a \PQb quark and a \PW boson before hadronization, the spin information is propagated to the \PW boson and thus to the angular distribution of the leptons. For \tDM \tW-channel processes, the (pseudo)scalar mediator is radiated from the top quark, affecting the final top polarization and therefore the correlation with the \PW boson. This observable contains information regarding the polarization correlation of the two particles and helps in separating the \tWDM from the SM single top quark \tW production because of the effect of the mediator coupling to top quarks.

\item $\abs{\Delta\phi(\ptvecmiss,\Pell\PAell\PQb)}$, the opening angle between the two-lepton plus (leading) \PQb-tagged jet and \ptvecmiss. This variable allows differentiation of \tDM events from the dominant \ttbar background since in \tDM events the $\Pell\PAell\PQb$ system is expected to be back-to-back with respect to \ptvecmiss. This is not the case for the \ttbar production, which has another \cPqb jet present. 
\end{itemize}

In the 2 \PQb tag region, the four-momenta of the top quarks can be reconstructed from all the visible decay products of the \ttbar system. This can be achieved by constraining the neutrino four-momenta using the on-shell conditions for the top quarks, \PW bosons, and neutrinos along with the fact that the \ptmiss for \ttbar processes should correspond to the sum of the transverse momentum of the neutrinos. While this condition is valid for \ttbar processes, it does not hold for the \ttDM signal where the DM particles produced also contribute to the total \ptmiss in the event. Therefore, a variant of traditional kinematic reconstruction algorithms~\cite{Betchart:2013nba} is used. This algorithm first applies the on-shell conditions to limit possible neutrino solutions to ellipses in momentum space and then attempts to assign the neutrino momenta as closely as possible to \ptmiss. If the entire \ptmiss can be assigned to neutrino momenta, the ellipses for the neutrino and anti-neutrino will intersect, and the point of intersection with the smallest $m_{\ttbar}$ is taken as a solution for the neutrino momenta. Otherwise (as is usually the case because of the $\mTiill > 80\GeV$ selection), the ellipses do not intersect, and the point of closest approach is taken as a solution. Events that fail this kinematic reconstruction (which only happens if the on-shell conditions are not satisfied and represents about 5\% of signal and background events) are not considered in this region. This reconstruction of the top quarks allows the use of additional discriminating variables in the 2 \PQb tag region based on the top quark kinematic properties:

\begin{itemize}
\item $p^{\text{dark}}_{\text{T}}$: this variable defines the amount of \ptmiss that cannot be assigned to the neutrinos in the top quark kinematic reconstruction, \ie, the distance of closest approach of the ellipses. This variable tends to have higher values for \ttDM events with respect to \ttbar production because of the additional source of genuine \ptmiss from DM particles.
\item $c_{\text{hel}}$: the angle between the two leptons when boosted back into the rest frame of their parent top quarks. This quantity is sensitive to spin correlations. 
\item $\abs{\Delta\phi(\PQt,\PAQt)}$: the top quark and antiquark are expected to be approximately back-to-back in the azimuthal plane for \ttbar production, while for \ttDM events they are expected to recoil against the DM mediator.
\end{itemize}

The NNs were trained using Keras with a Tensorflow backend. Each contained three densely connected hidden layers using Recified Linear Unit (ReLU) activation function, and an output layer with nodes for signal and background using the ``softmax'' activation function. Categorical cross-entropy was used for the activation function, and the network was trained for 50 epochs. The MC data sets were split into training, validation and test data sets; no evidence was observed for overtraining between the training and validation data sets, and no significant difference in the final distributions was observed between the test and training data sets. 
The NN score was used in the signal extraction, explained in more detail in Section~\ref{sec:fit}, with a binning designed to maximize sensitivity while keeping a minimum of 5 expected background events in each bin.

\begin{table}[htbp]
\centering
\topcaption{Final event selection requirements for the AH, SL, and DL SRs. For the SL channel, a categorization in terms of modified topness, with bins of $t\leq 0$ and $t>0$, is also applied after the event selection. The DL channel is split into SF $\Pe^{+}\Pe^{-}$/$\PGm^{+}\PGm^{-}$ and DF $\Pe^{\pm}\PGm^{\mp}$ regions.}
\cmsTable{
\begin{tabular}{lccc{c}@{\hspace*{5pt}}ccc{c}@{\hspace*{5pt}}cc}
\hline
& \multicolumn{ 3}{c}{All-hadronic SRs} & & \multicolumn{ 3}{c}{Single-lepton SRs} & & \multicolumn{ 2}{c}{Dilepton SRs} \\\cline{2-4}\cline{6-8}\cline{10-11}
& 0$\Pell$, 1 \cPqb, 0FJ & 0$\Pell$, 1 \cPqb, 1FJ & \hspace*{3mm} 0$\Pell$, 2 \cPqb \hspace*{3mm}& & 1$\Pell$, 1 \cPqb, 0FJ & 1$\Pell$, 1 \cPqb, 1FJ & \hspace*{2mm} 1$\Pell$, 2 \cPqb \hspace*{2mm}& & 2$\Pell$, 1 \cPqb & 2$\Pell$, 2 \cPqb\\
\hline
\nlep        & \multicolumn{ 3}{c}{$=$0}     & & \multicolumn{ 3}{c}{$=$1}     & & \multicolumn{ 2}{c}{$=$2} \\
\njet        & \multicolumn{ 3}{c}{${\geq}3$} & & \multicolumn{ 3}{c}{${\geq}2$} & & \multicolumn{ 2}{c}{${\geq}1$} \\
\nbjet       & $=$1  & $=$1      & ${\geq}2$  & & $=$1 & $=$1       & ${\geq}2$ & & $=$1 & ${\geq}2$ \\
Forward jets & $=$0  & ${\geq}1$ & \NA        & & $=$0 & ${\geq}1$  & \NA       & & \NA   & \NA       \\
\HTratio     & \NA & \NA & $<$0.5      & &  \multicolumn{ 3}{c}{\NA}                & &  \multicolumn{ 2}{c}{\NA} \\
\ptmiss [\GeVns{}]     & \multicolumn{ 3}{c}{$>$250}        & &  \multicolumn{ 3}{c}{$>$250}         & &  \multicolumn{ 2}{c}{\NA} \\
\mT [\GeVns{}]          & \multicolumn{ 3}{c}{\NA}               & &  \multicolumn{ 3}{c}{$>$140}         & &  \multicolumn{ 2}{c}{\NA} \\
\mTt [\GeVns{}]         & \multicolumn{ 3}{c}{\NA}               & &  \multicolumn{ 3}{c}{$>$180}         & &  \multicolumn{ 2}{c}{\NA} \\
\mindphi [rad.]     & \multicolumn{ 3}{c}{$>$0.8} & &  \multicolumn{ 3}{c}{$>$0.8}  & &  \multicolumn{ 2}{c}{\NA} \\
\mTb [\GeVns{}]         & \multicolumn{ 3}{c}{$>$140}        & &  \multicolumn{ 3}{c}{$>$140}         & &  \multicolumn{ 2}{c}{\NA} \\
\mll [\GeVns{}]       & \multicolumn{ 3}{c}{\NA}                & &  \multicolumn{ 3}{c}{\NA}                & &  \multicolumn{ 2}{c}{$>$20} \\
$\abs{\mll-m_{\PZ}}$ [\GeVns{}] & \multicolumn{ 3}{c}{\NA}           & &  \multicolumn{ 3}{c}{\NA}                & &  \multicolumn{ 2}{c}{$>$15 (SF)} \\
\mTiill [\GeVns{}]        & \multicolumn{ 3}{c}{\NA}             & &  \multicolumn{ 3}{c}{\NA}                & &  \multicolumn{ 2}{c}{$>$80} \\
Pass \ttbar reco & \multicolumn{ 3}{c}{\NA}           & &  \multicolumn{ 3}{c}{\NA}                & & \NA & yes \\
\hline
\end{tabular}
}
\label{tab:SR}
\end{table}

\subsection{Control regions}

After the event selection presented in Table~\ref{tab:SR} is applied, the leading SM background contributions in the different regions are evaluated. For the AH regions, the main backgrounds arise from single-lepton \ttbar and \wjets events in which the lepton is not identified, and {\cPZ} boson production in which the {\cPZ} boson decays into two neutrinos, leading to genuine \ptmiss. In the SL SRs, the main backgrounds are dileptonic \ttbar events in which one of the leptons is not identified, and \wjets events. In the DL SRs, the backgrounds are dileptonic \ttbar and \tW events, DY production, and \ttZtonunu processes.

To improve the estimation of the backgrounds, control samples in data are used. In the AH and SL channels, CRs enhanced in the different background sources are used to derive correction factors as a function of \ptmiss from the comparison of the \ptmiss distribution between data and simulation. These corrections are extracted simultaneously across the CRs and SRs for each channel in a global fit, as explained in more detail in Section~\ref{sec:fit}. The residual background processes are modeled with simulation.

In the DL channel, a validation region (VR) enriched in \ttbar and \tW events is used to verify their modeling for all the input variables used in the NN. These distributions all show good agreement between data and simulations within uncertainties and therefore no correction factor is used. For the DY and \ttZtonunu processes, dedicated CRs are used to derive corrections in the fit. For the DY process, these corrections are derived as a function of the NN score in a similar way to the AH and SL channels, while for \ttZtonunu, only an overall normalization factor is considered.

\subsubsection{All-hadronic control regions}
For the AH SRs, four independent sets of CRs are defined. The first set of CRs is designed to isolate single-lepton \ttbar processes by selecting events with exactly one lepton, $\njet\geq3$, $\nbjet\geq1$, $\ptmiss>250\GeV$, $\mindphi\geq 0.8$, and $\mT<140\GeV$ to avoid overlap with the SL SRs.

The second set of CRs is enhanced in single-lepton \wjets processes by selecting events with exactly one lepton, $\njet\geq3$, $\nbjet=0$, $\ptmiss>250\GeV$, and in order to avoid overlap with the SL \wjets CR, $\mT<140\GeV$.

Additional CRs are designed to model the background originating from \zjets production, where the {\PZ} boson decays into a pair of neutrinos (\Zinv). Here we use the {\PZ} boson decays to an OS SF dilepton pair (\ZLL) as proxy events to emulate the kinematic properties of the \Zinv process. Events are selected by requiring two leptons with SF (\ie, $\Pe^{+}\Pe^{-}$ or $\PGm^{+}\PGm^{-}$) and OS that satisfy the requirement $60<\mll<120\GeV$ on their invariant mass. Additionally, events must contain at least three jets and events with {\cPqb}-tagged jets are vetoed ($\nbjet=0$). To reproduce the \pt spectrum of \Zinv events, the two leptons are added to the \ptvecmiss, giving a quantity referred to as hadronic recoil, which is required to be greater than 250\GeV.

The fourth CR is introduced to estimate QCD multijet events from data to reduce their associated uncertainties. Though QCD multijet events are a minor background in the AH SRs, they are characterised by large systematic uncertainties. These large systematic uncertainties are due to the fact that to pass the $\ptmiss>250\GeV$ selection, they need to have a significant jet mismeasurement, which is difficult to model accurately in simulation. The associated CR is defined by requiring zero leptons, $\njet\geq3$, $\nbjet\geq1$,  $\ptmiss>250\GeV$, and $\mindphi< 0.8$. The last requirement selects events where the \ptmiss is aligned with one of the leading jets, pointing to a mismeasurement.

A summary of the different AH CRs can be found in the first four columns of Table~\ref{tab:CR}.

\subsubsection{Single-lepton control regions}
\label{sec:SLCR}
The first set of CRs is designed to estimate dileptonic \ttbar events by requiring exactly two leptons, $\njet\geq2$, $\nbjet\geq1$, and $\ptmiss>250\GeV$. Additionally, events are required to have $\mTiill<80\GeV$ to maintain the orthogonality with the DL SRs. To statistically enhance these CRs, the \mT, \mTt, and forward jet selections are removed.

The second set of CRs is defined to isolate \wjets events by requiring exactly one lepton, $\njet\geq2$, $\nbjet=0$, $\ptmiss>250\GeV$, and $\mT>140\GeV$. The $\nbjet=0$ requirement makes this CR orthogonal to the SL SRs.

The selection criteria are summarized in the central columns of Table~\ref{tab:CR}.

\subsubsection{Dilepton control regions}
\label{sec:DLCR}
For the DL channel, a VR is used to check the agreement for the dominant \ttbar background between data and simulation for the distributions used as input variables to the NN. This region is defined with the same selections used in the DL SRs (exactly two leptons, at least one jet and {\PQb}-tagged jet, \mll$>20$\GeV, and outside the \PZ boson mass window for SF leptons), except for the inversion of the \mTiill selection requirement. No further correction to these processes in the fit is found to be necessary.

The DY process is found to be not well modeled in simulation. In particular, the \ptmiss distribution and the \mTiill distribution, which are the two most sensitive NN input variables, are not well described. A CR is defined with an identical selection to the DL SRs (exactly two leptons, at least one jet and {\PQb}-tagged jet, \mll$>20$\GeV, and \mTiill$>$80\GeV), except only SF ($\Pe^{+}\Pe^{-}$/$\PGm^{+}\PGm^{-}$) leptons are considered inside the $\PZ$ boson mass window $76<\mll<106$\GeV. Separate CRs corresponding to the 1 \PQb tag and 2 \PQb tag regions were considered, since this distribution is also known to be poorly modeled. However, the 2 \PQb tag region was found to not be dominated by DY events even inside the \PZ boson mass window. Therefore, a single CR requiring exactly 1 \PQb-tagged jet is used to predict the rate of DY in the 1 \PQb tag SF SR as a function of the NN score.

To model the background originating from \ttZtonunu production, a CR enriched in \ttZ processes, where the \PZ decays into electrons or muons, is used. 
To reduce the statistical uncertainty, a CR targeting SL \ttbar decays is used, which is defined by requiring exactly three leptons, with $\pt>25, 20, 20\GeV$ respectively, along with at least three jets with $\pt > 30\GeV$ for the leading jet, $\pt > 20\GeV$ for all others, and at least one {\PQb}-tagged jet. Furthermore, events are required to have OS SF lepton pair with $\abs{\mll - m_{\PZ}} < 10\GeV$, and the remaining lepton to have $\pt>35\GeV$ to suppress processes coming from DY with a misidentified lepton. To maximize the discrimination of \ttZ from the remaining backgrounds, the events are binned in terms of \njet and \nbjet, since \ttZ has higher multiplicities of both compared to the other backgrounds.

A summary of the DL VR and CRs can be found in the last three columns of Table~\ref{tab:CR}.

\begin{table}[htbp]
    \centering
    \topcaption{CRs defined for the main backgrounds of the AH SRs (first 4 columns, \ttonel, \wjets, \ZLL, QCD), the SL SRs (central two columns, \tttwol and \wjets), and the DL SRs (last 2 columns, \tttwol and \ttZ). To increase the event counts, some selection criteria applied in the SRs are removed in the related CRs and hence are not listed. The \ptmiss selection for the \ZLL CR refers to the hadronic recoil.}
    \cmsTable{
    \begin{tabular}{lccc{c}@{\hspace*{5pt}}cc{c}@{\hspace*{5pt}}cccc}
    \hline
     & \multicolumn{4}{c}{AH} & & \multicolumn{2}{c}{SL} & &  \multicolumn{3}{c}{DL} \\ \cline{2-5}\cline{7-8}\cline{10-12}
    & \ttonel CR & $\PW(\Pell\PGn)$ CR & \Ztwol CR & QCD CR & & \tttwol CR & $\PW(\Pell\PGn)$ CR   & & \tttwol VR & DY CR & \ttZ CR\\ 
    \hline
    $n_{\PQb}$            & $\geq$1      & $=$0          & $=$0                      & $\geq$1  & & $\geq$1         & $=$0    & &  $\geq$1                    & $=$1                 & $\geq$1\\
    $n_{\text{lep}}$       & $=$1          & $=$1          & $=$2                       & $=$0        &  & $=$2  & $=$1    & &  $=$2                       & $=$2 (SF)     & $=$3 \\
    $n_{\text{jet}}$       & $\geq$3       & $\geq$3       & $\geq$3                    & $\geq$3   & & $\geq$2         & $\geq$2        & &  $\geq 1$                    & $\geq$1                 & $\geq$3\\ 
    \ptmiss [\GeVns{}]      & $\geq$250 & $\geq$250 & $\geq$250 & $\geq$250     & & $\geq$250   & $\geq$250    & &  \NA                         &  \NA                     & \NA    \\ 
    $M_{\text{T}}$   [\GeVns{}]  & $\leq$140 & $\leq$140 & \NA                         & \NA         & & \NA              & $\geq$140    & &  \NA                         &  \NA                     & \NA    \\
    \mindphi [rad.]           &$\geq$0.8     & \NA            & \NA                         & $<$0.8        & & \NA              & \NA          & &  \NA                         &  \NA                     & \NA    \\
    \mll     [\GeVns{}]            & \NA            & \NA            & $[60, 120]$             & \NA         & & \NA              & \NA         & &  $>$20                   &  \NA                     & \NA   \\
    $\abs{\mll-m_{\PZ}}$  [\GeVns{}]   & \NA            & \NA            & \NA                         & \NA        & & \NA              & \NA       & &  $>$15 (SF)    & $<$15 & $<$10 (OS SF)\\
    \mTiill       [\GeVns{}]      &\NA            & \NA            & \NA                         & \NA        & & $\leq$80   & $\leq$80        & &  $\leq$80              & $\geq$80           & \NA \\ 
    Included in fit?    & Yes            & Yes            & Yes                         & Yes             & & Yes              & Yes         & &  No                          & Yes                       & Yes    \\ 
    \hline
    \end{tabular}
    }
    \label{tab:CR}
\end{table}

\section{Systematic uncertainties}
\label{sec:systematics}
Different sources of systematic uncertainties are considered, in most cases affecting both the signal and the background. The various sources are implemented in the limit calculation, described in detail later. We distinguish between two types of uncertainties, those that only affect the normalization of a process, and those that in addition affect its distribution.

The following uncertainties correspond to constrained normalization nuisance parameters in the fit discussed in Section~\ref{sec:fit} (unless otherwise noted, the uncertainty source is applied to all channels):

\begin{itemize}
\item \textit{Luminosity}: 
the integrated luminosities for the 2016, 2017, and 2018 data-taking years have 1.2--2.5\% individual uncertainties~\cite{CMS-LUM-17-003,CMS-PAS-LUM-17-004,CMS-PAS-LUM-18-002}.

\item \textit{Single top quark background normalization (AH and SL channels)}:
an uncertainty of 20\% in the normalization of single top quark processes is considered, accounting for the uncertainty due to differences in generator predictions and the \ttbar/\tW interference treatment in the predicted cross section.
\end{itemize}

The following sources of uncertainty affect the distributions as well as the normalization of the signal and background processes. They are applied to all search channels:

\begin{itemize}
\item \textit{PDF uncertainties}: 
uncertainties in the choice of PDF are estimated by reweighting the samples to the NNPDF~3.0~\cite{Ball:2014uwa} replicas for the 2016 data set and the NNPDF~3.1~\cite{NNPDF:2017mvq} hessian variations for the 2017 and 2018 data sets, which were then combined as described in Ref.~\cite{Butterworth:2015oua}.

\item \textit{Factorization and renormalization scales}:
 the uncertainties in the choice of the factorization and renormalization scale parameters are taken into account by applying a set of weights that represent a change of these scales by a factor of 2 or 0.5. For processes where the rate is estimated directly from data in a CR (such as \PW+jets, \PZ+jets, and QCD), these uncertainties are not considered;

\item \textit{Parton shower modeling}: 
variations in the parton shower initial and final state radiation (ISR and FSR) scales are also considered by independently varying these up and down by a factor of 2; 

\item \textit{Pileup modeling}:
the systematic uncertainty in the pileup modeling is taken into account by varying the total inelastic cross section entering the calculation of pileup distributions in simulation by $\pm4.6\%$~\cite{Sirunyan:2018nqx};

\item \textit{Trigger}: 
separate uncertainties are considered for the triggers employed in each channel. These are estimated using cross-checks of the methods used to compute them from control samples in data; 

\item \textit{Lepton reconstruction and selection}: 
scale factors are applied to the MC processes to mimic the measured reconstruction and selection efficiencies of leptons in data. The uncertainties associated with these scale factors are binned in \pt and $\eta$ and are of the order of $\sim$2\% for electrons and muons;

\item \textit{Jet energy scale and resolution}: 
reconstructed jet four-momenta in the simulation are varied according to the uncertainties in the jet energy scale, which are split into different uncorrelated sources. Additionally, the \pt of the jets is stochastically varied within the resolution of the detector. These uncertainties are coherently propagated to all variables, including \ptmiss~\cite{Khachatryan:2016kdb};

\item \ptmiss \textit{mismodeling}: 
the effects of varying jet uncertainties are propagated to the calculation of the \ptmiss. Additionally, any unclustered energy, mostly coming from calorimeter deposits not assigned to any object, is varied within the associated measurement uncertainties and the impact of such variations is propagated to \ptmiss;

\item \textit{{\PQb} tagging efficiency scale factors}: 
the {\PQb} tagging and light-flavor quark/gluon jet mistag efficiency scale factors and the respective uncertainties are measured in independent control samples~\cite{CMS:2017wtu} and propagated to the analysis;

\item \textit{\PW\unskip/\PZ\unskip+heavy-flavor fraction (AH and SL channels)}:
the uncertainty in the fraction of \PW\unskip/\PZ\unskip+heavy-flavor (HF) jets in \wjets and \zjets events is taken into account. The fraction of \PW\unskip/\PZ\unskip+jets events in which the jets include heavy flavour jets is varied within 20\% separately and independently for the one \PQb-tagged and two or more \PQb-tagged jets cases~\cite{CMS:2016sun,CMS:2016eha,CMS:2014bkk,CMS:2016gmz}.

\item \textit{Simulation sample size}: 
uncertainties originating from the limited size of the simulated signal and background samples are included by allowing each kinematic bin used in the signal extraction to fluctuate independently within its statistical uncertainty following Ref.~\cite{BARLOW1993219};

\item \textit{Uncertainty related to ECAL mistiming}:
partial mistiming of signals in the forward regions of the ECAL endcaps led to a minor reduction in trigger efficiency. Simulations are corrected to reproduce the behavior of the data and the uncertainty in these corrections is propagated to the distributions used in the signal extraction;

\item \textit{Electroweak and QCD $K$ factors (AH and SL channels)}: 
  uncertainties in the LO$\to$NLO $K$ factors calculated for \wjets and \zjets processes are included to compensate for missing higher-order corrections. For QCD processes, the uncertainties arise from variations related to the factorization and renormalization scales. For electroweak processes, the magnitude of the missing higher-order corrections is estimated by calculating the difference between applying and omitting the LO$\to$NLO electroweak $K$ factors.

\item \textit{{\PQb}-tagged jet multiplicity normalization (AH and SL channels)}: 
an uncertainty of 5\% is considered for the normalization scale factors applied to both \wjets and \zjets processes in 2017 and 2018 to correct the \PQb-tagged jet multiplicity in the AH and SL channels;

\item \textit{Top quark \pt reweighting (DL channel)}: 
differential measurements of the top quark \pt spectrum in $\ttbar$ events~\cite{Khachatryan:2016mnb} show that the measured \pt spectrum is softer than in NLO simulation. Comparisons with NNLO QCD simulation suggest this difference is mainly due to missing higher-order corrections. A \pt-dependent scale factor is therefore applied to correct the difference between NNLO and NLO simulation in the DL channel. This correction is not needed in the other channels because of the \ptmiss-dependent CR corrections factors. A corresponding systematic uncertainty is estimated by evaluating the difference between applying and omitting the reweighting.
\end{itemize}

The above sources of systematic uncertainties are considered with various degrees of correlation across the different channels and the data-taking periods. A group of nuisance parameters, such as the theory-related uncertainties, which include the variation in the factorization and renormalization scales, the PDF choice, the parton showering scales, the uncertainty in the inelastic cross section, and others, are treated as fully correlated across all channels and periods of data-taking. Analogously, for the channel-dependent higher-order theory corrections to \wjets, \zjets, and $\ttbar$ processes, a commonly associated parameter is assigned across all periods. Experimental sources of uncertainties such as the lepton identification and isolation efficiencies, as well as the efficiency of the different triggers used, are also considered fully correlated. Similarly, a subgroup of size equal to 13 nuisance parameters out of the 27 sub-sources affecting the jet energy scale and the parameter associated with the ECAL mistiming uncertainty are assumed to be maximally correlated across all analysis categories. Another group of nuisance parameters is considered in the fit with a partial correlation scheme. This includes, for instance, uncertainty sources such as b tagging efficiency, for which a fraction of sub-sources are considered fully correlated and another fraction completely uncorrelated across data-taking periods. The remainder (14 parameters) of the group of jet energy scale sub-sources are also treated as fully correlated, making the jet energy scale a partially correlated uncertainty source overall. The uncertainty in the integrated luminosity is naturally correlated across channels but is kept uncorrelated among different data-taking periods. On the other side, among the group of fully uncorrelated uncertainty sources across periods, one finds the nuisance parameters associated with jet energy resolution and the unclustered component of the missing transverse momentum. These two, however, because of their common nature, are kept correlated across the three channels investigated here.

\section{Signal extraction}
\label{sec:fit}

The DM signal is extracted from a simultaneous fit to the \ptmiss distributions in the AH and SL channels, and to the NN output distributions for the DL channel.
The following results have been determined using the CMS statistical analysis tool \textsc{Combine}~\cite{CAT-23-001}, which is based on the \textsc{RooFit}~\cite{Verkerke:2003ir} and \textsc{RooStats}~\cite{Moneta:2010pm} frameworks.
The main SM backgrounds, discussed in Section~\ref{sec:Selection}, are dileptonic \ttjets and \wjets events for the SL channels, single-lepton \ttjets, \wjets, and \Zinv events for the AH channels, and dileptonic \ttjets, \tW, and \ttZtonunu for the DL case.

The effect of the systematic uncertainties is taken into account by introducing nuisance parameters. Uncertainties that affect the normalization only are modeled using nuisances with log-normal probability densities. These parameters are treated as correlated between bins of the fit distribution and between the different CRs and SRs within each channel for the same year. The common sources are correlated across channels, with the exception of the {\PQb} tagging efficiency scale factors, which are decorrelated between the AH, SL, and DL channels because of a significantly different phase space.

To improve the estimation of the primary backgrounds in the SL and AH channels, an unconstrained multiplicative parameter is separately assigned to each background for each individual bin of the \ptmiss distribution. A thorough explanation of this method, typically referred to as the ``transfer factor'' method in many other DM searches, can be found in Ref.~\cite{CMS:2025448}. Concretely, the unconstrained parameters tend to adjust the expected contribution of the corresponding background process as a function of \ptmiss simultaneously in the SRs and CRs for a particular channel, taking advantage of the similar distributions of the fitting variables across different regions. For example, in any given \ptmiss bin of the SL channel, a single multiplicative parameter for \ttbar links the \ttbar background in the \ttbar enhanced 2$\Pell$ CR, the \wjets enhanced 1$\Pell$ CR, and the SRs. This has the advantage that the larger expected sample of \ttbar events in the enhanced 2$\Pell$ CR provides the primary, though not exclusive, control of the necessary corrections to that given process in the SR by comparing to the observation in the signal-depleted region. 
Similarly, a multiplicative parameter links the rates of the DY process in each NN bin between the DY CR and the DL 1 \PQb tag SF SR, while an overall normalization parameter is used to link the rate of \ttZ between the \ttZ CR and the DL SRs. The rates of processes that do not have a dedicated CR in the fit are taken from simulation. Potential contributions from DM signals are included in all CRs and SRs, and scaled by a signal strength modifier $\mu=\sigma/\sigma_{\text{th}}$, \ie the ratio between the measured and the theoretical cross sections.

\begin{figure}[!htbp]
\centering
\includegraphics[width=0.49\textwidth]{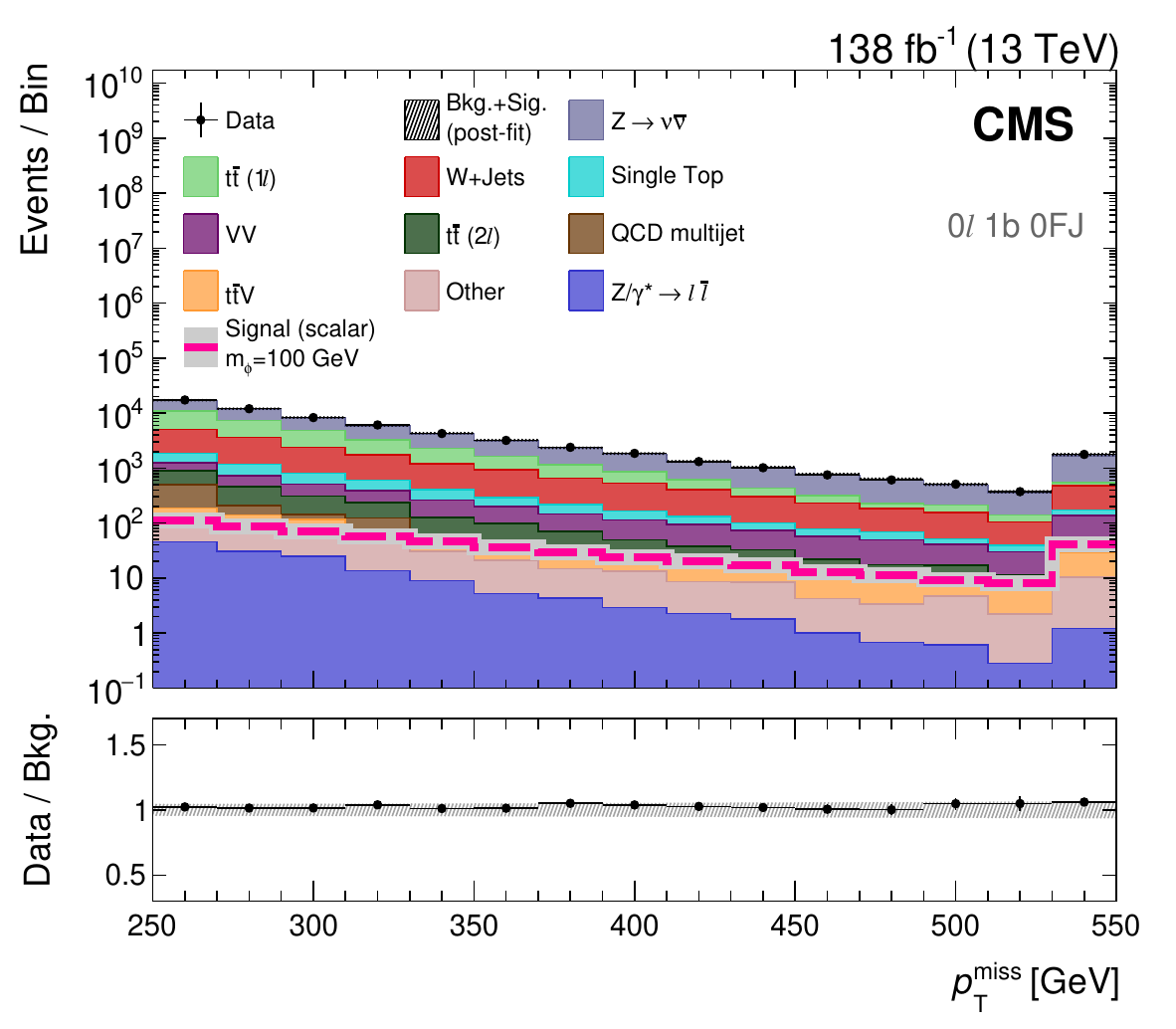}
\hfill
\includegraphics[width=0.49\textwidth]{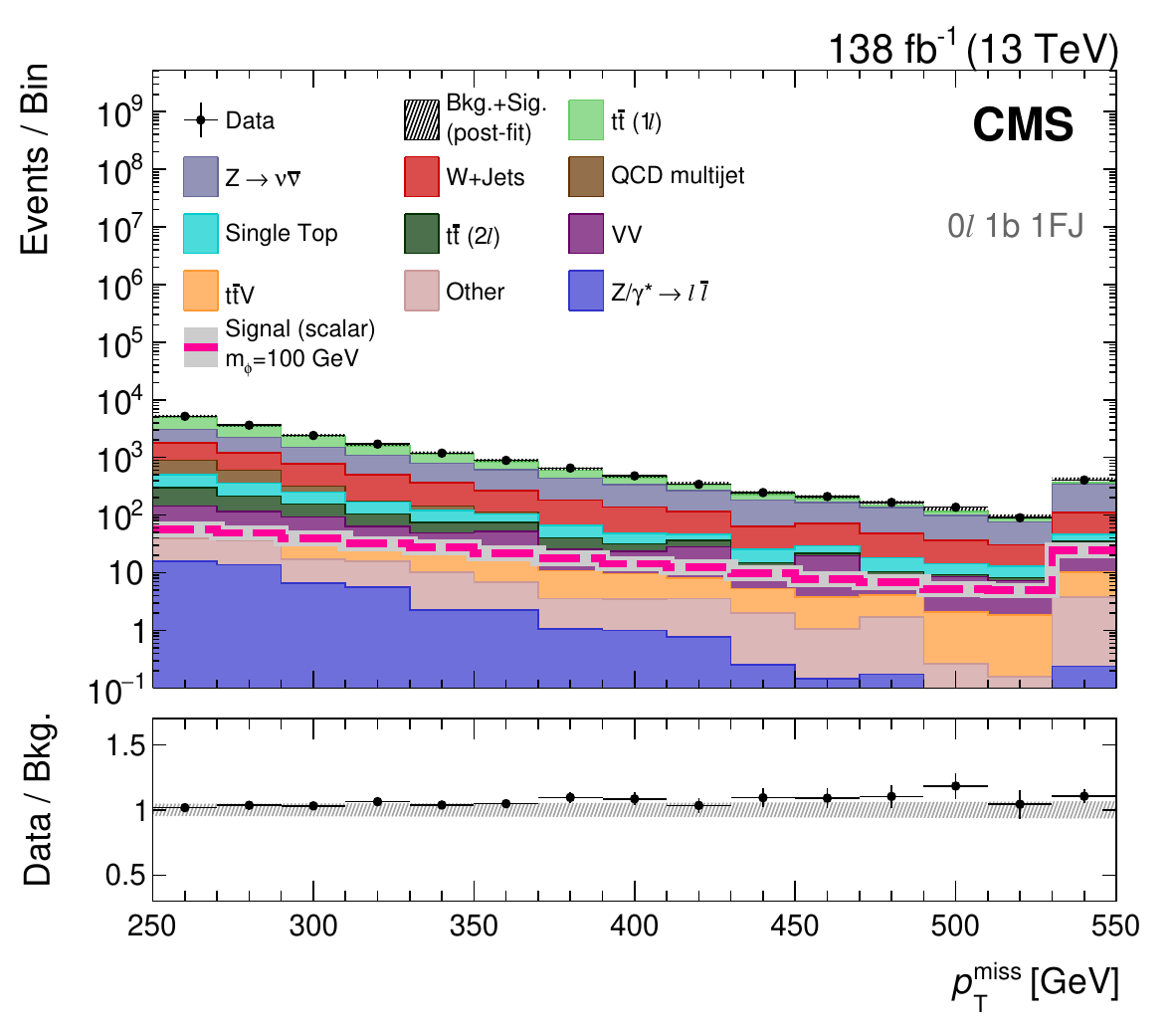}\\
\includegraphics[width=0.49\textwidth]{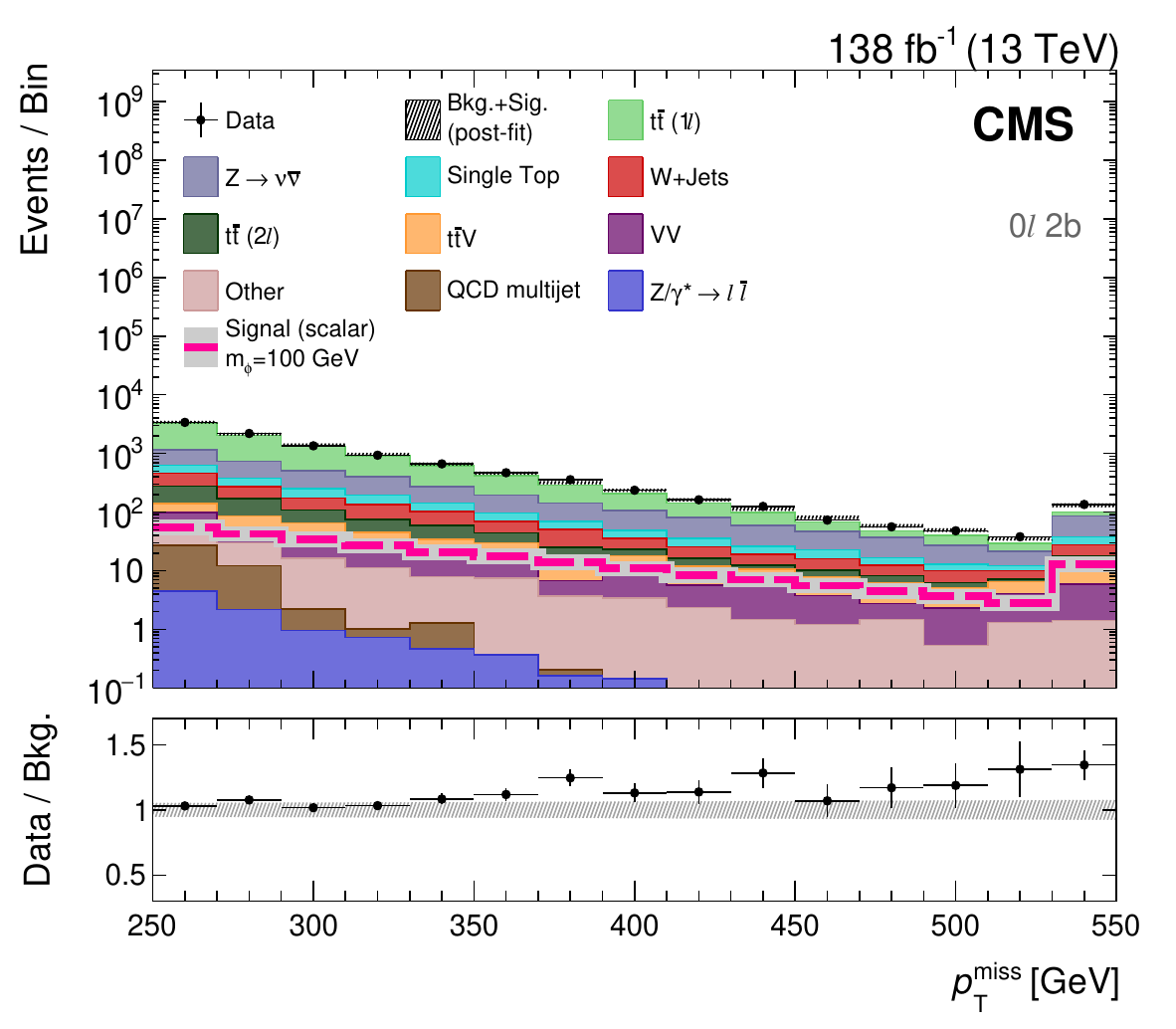}
\caption{The main discriminant distribution \ptmiss in the three AH SRs: 1b 0FJ (top left), 1b 1FJ (top right), and 2b (bottom), after a signal plus background fit across all channels, assuming a 100\GeV scalar mediator. The solid histograms for the simulated SM backgrounds are summed cumulatively and rescaled to luminosity, while the signal is shown as a dashed pink line. The grey dashed area in the upper panel represents the total uncertainty in the cumulative sum of the simulated SM backgrounds and signal, while in the lower panel it represents only the total uncertainty in the backgrounds. The data are represented by solid points with the horizontal bar indicating the width of the bin and the vertical one the associated statistical uncertainty. The last bin contains overflow events. The lower panel shows the ratio of data to post-fit background.}
\label{fig:ah:SRs:prepostfit}
\end{figure}

\begin{figure}[!htbp]
\centering
\includegraphics[width=0.45\textwidth]{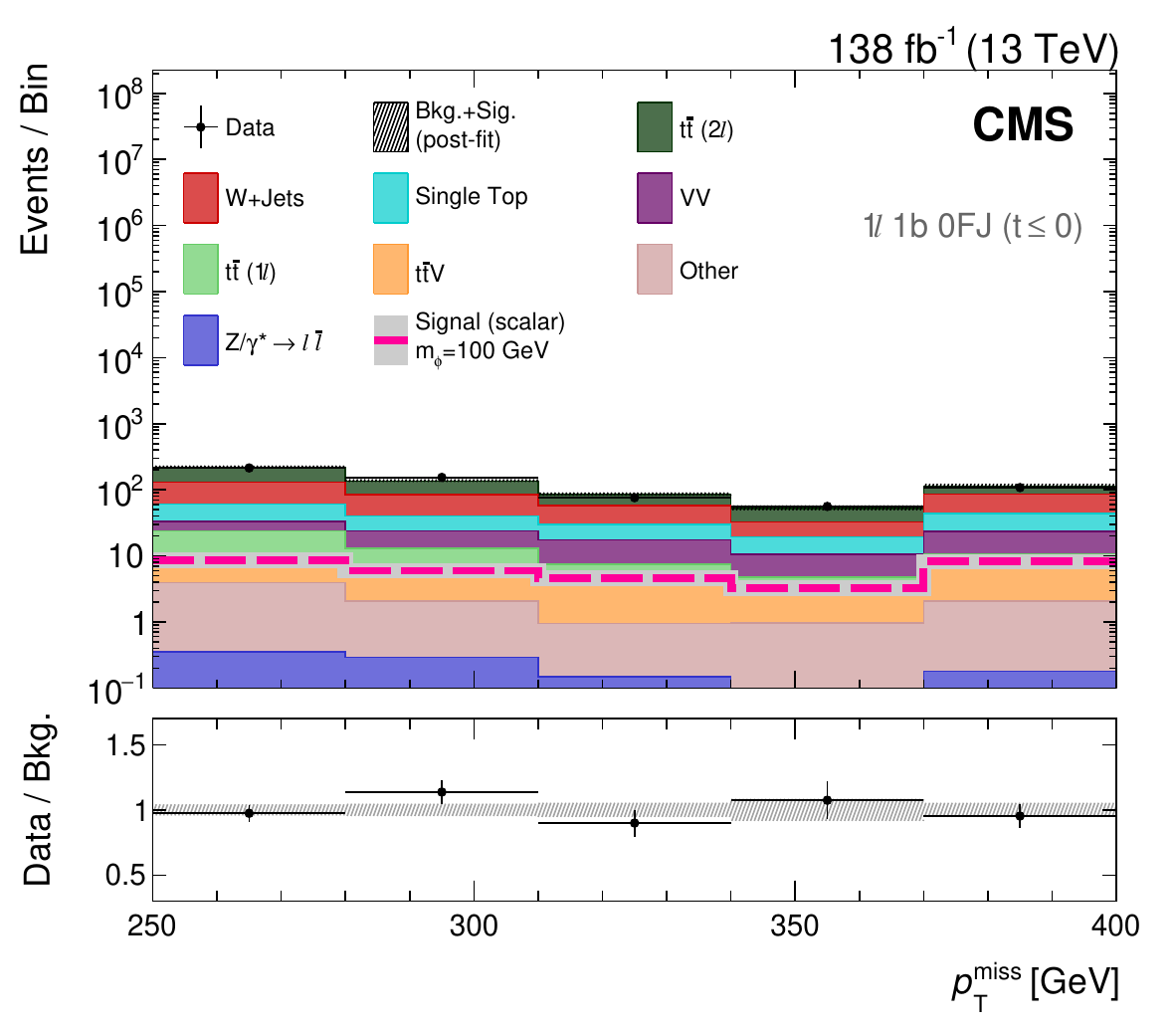}
\hfill
\includegraphics[width=0.45\textwidth]{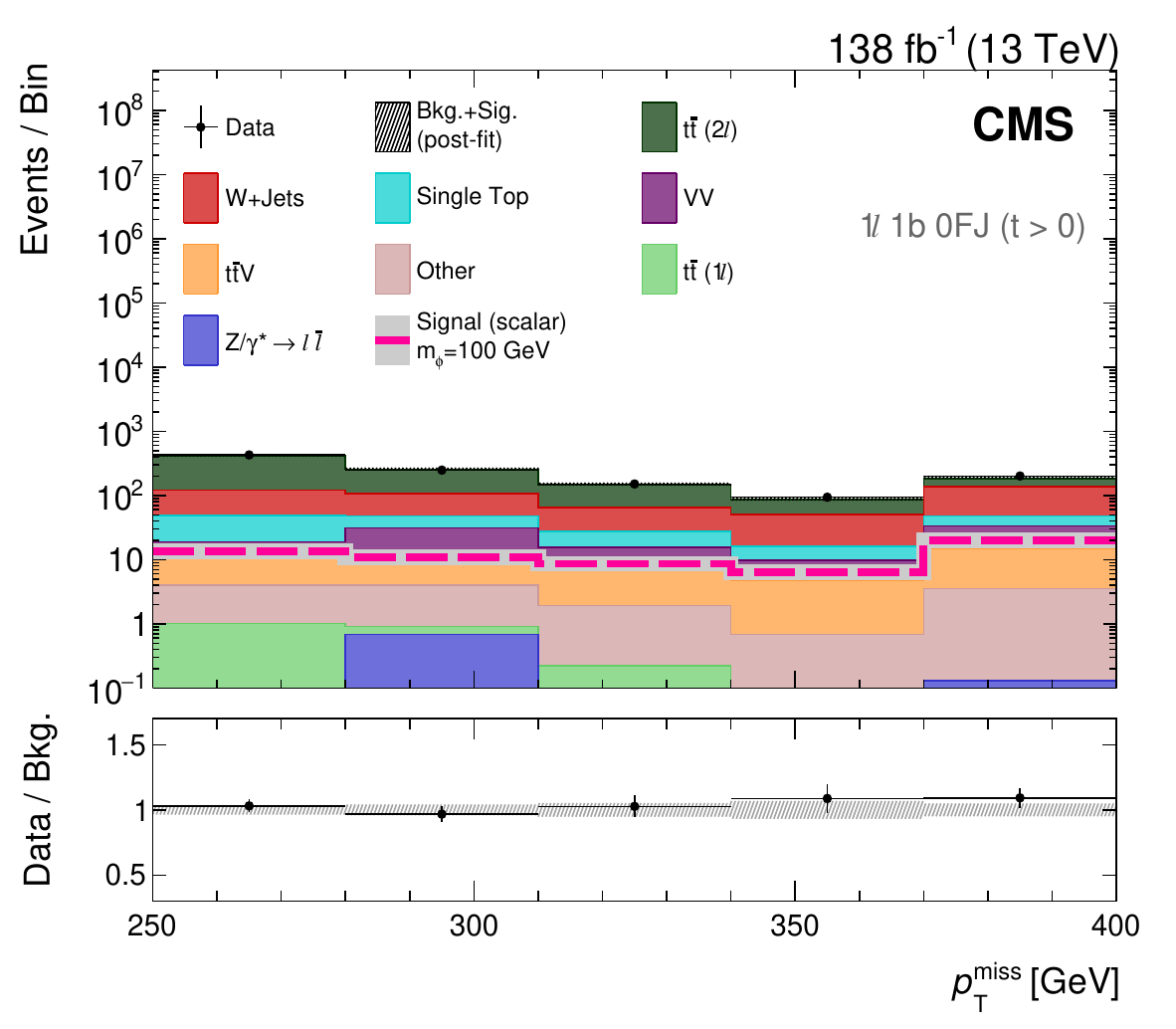}\\
\includegraphics[width=0.45\textwidth]{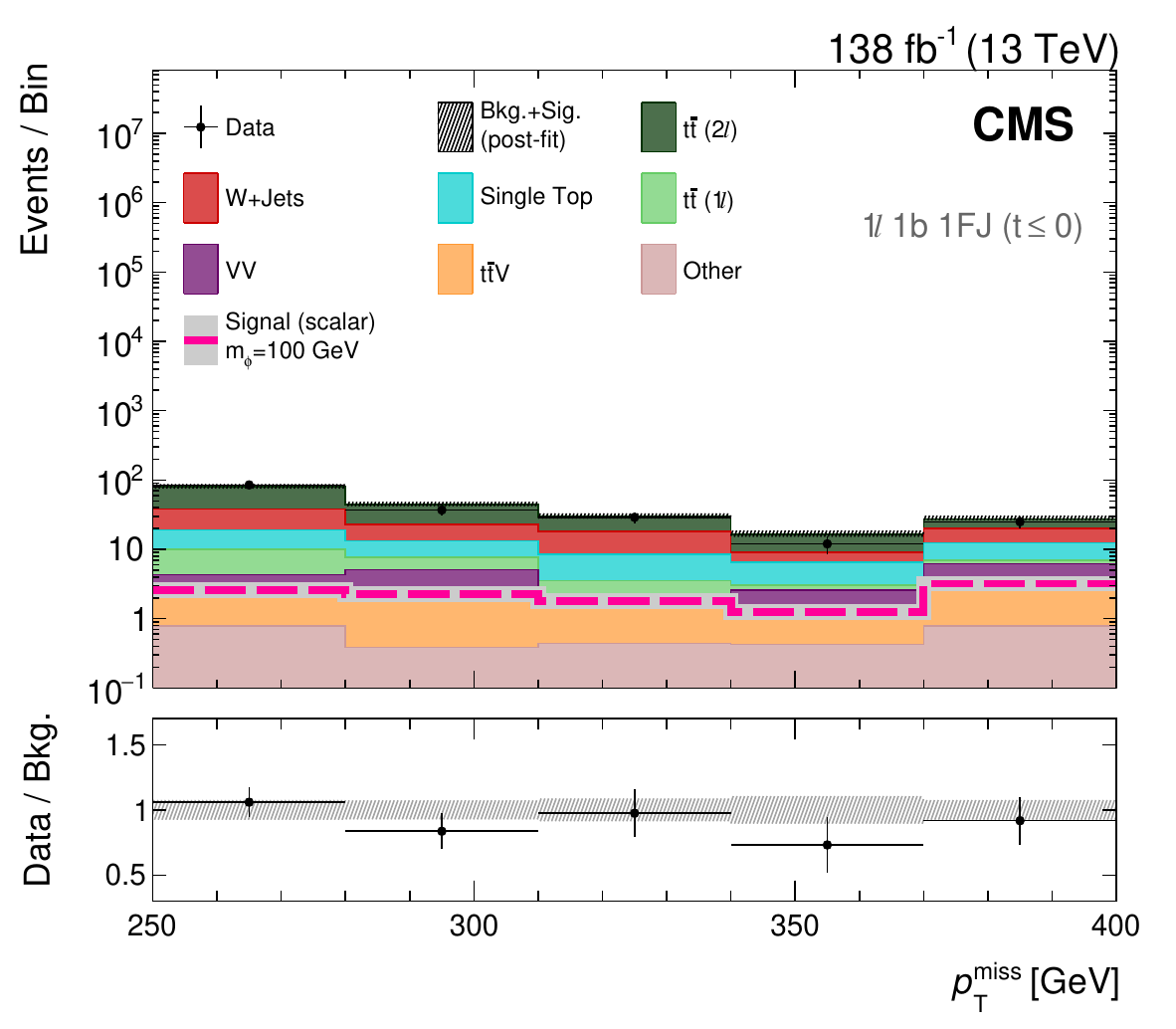} 
\hfill
\includegraphics[width=0.45\textwidth]{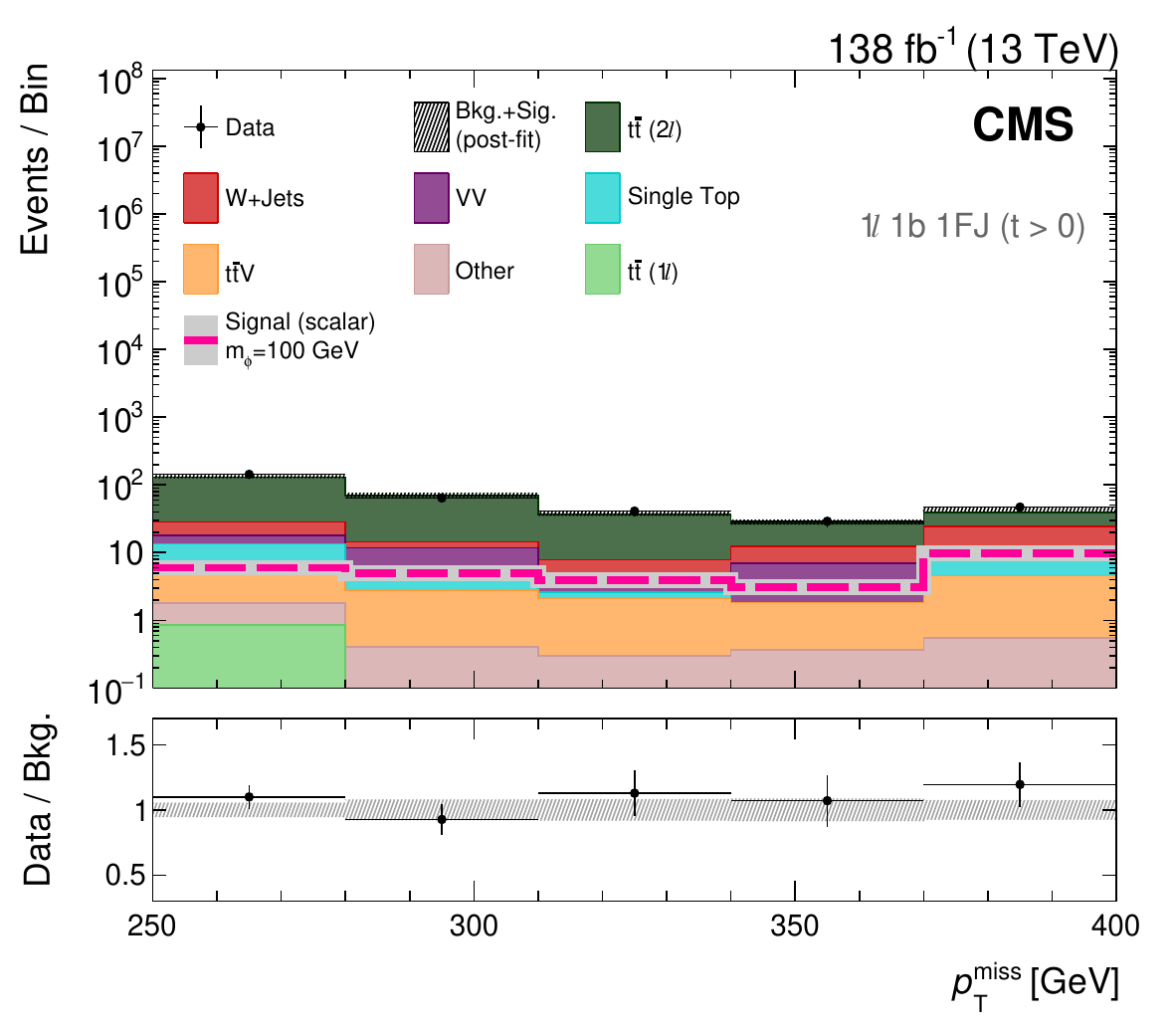}\\
\includegraphics[width=0.45\textwidth]{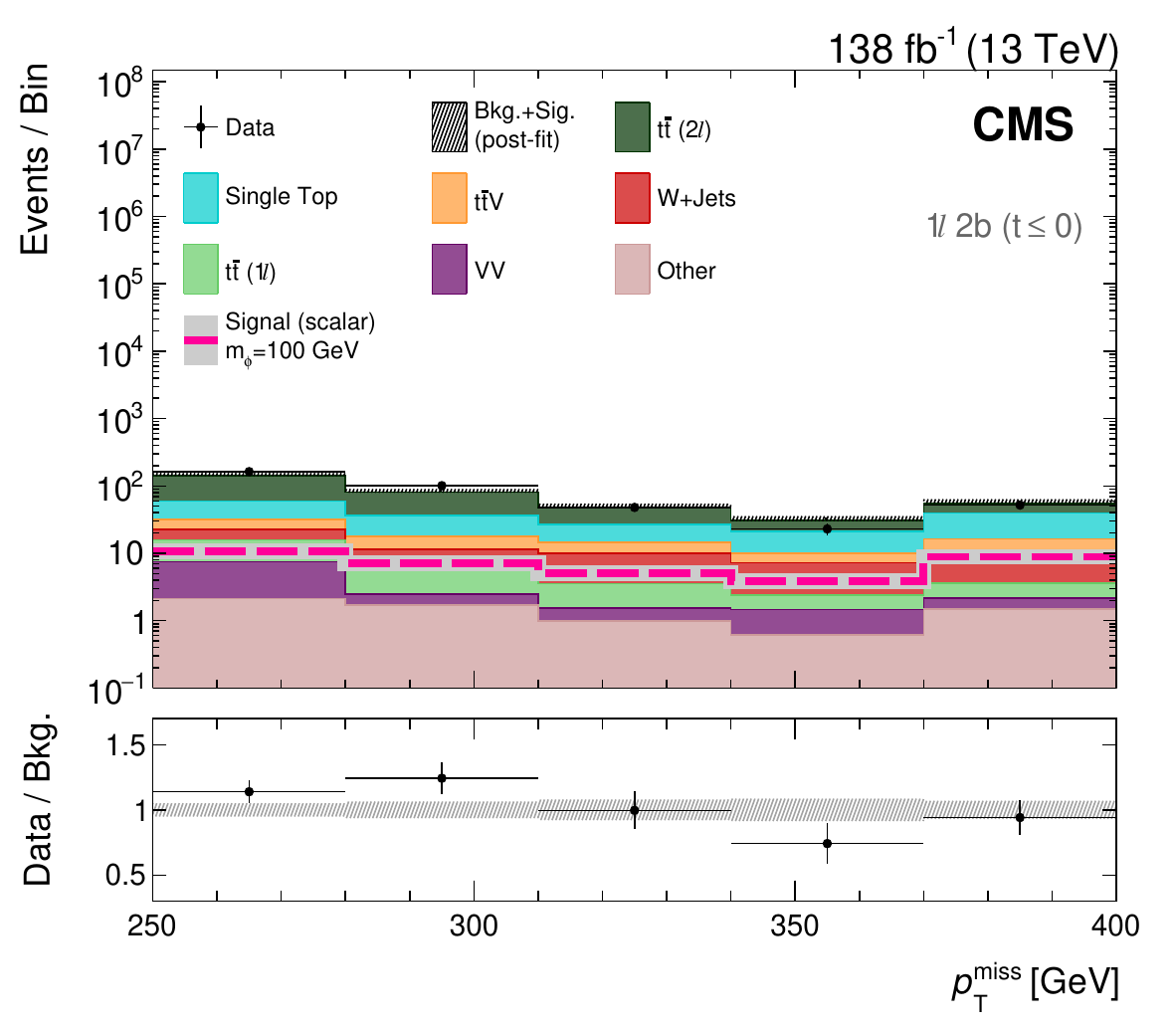}
\hfill
\includegraphics[width=0.45\textwidth]{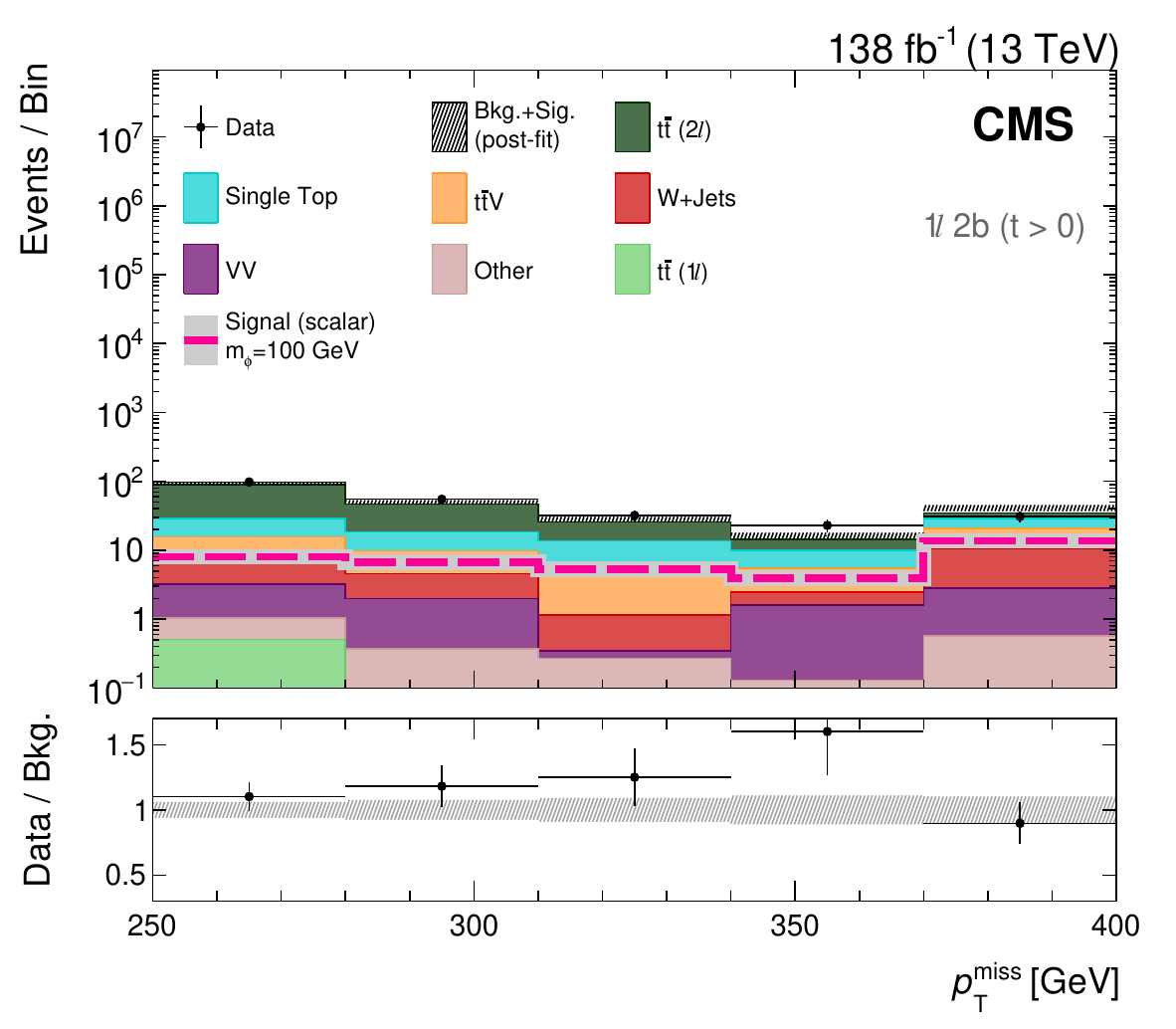}
\caption{The main discriminant distribution \ptmiss in the six SL SRs: 1b 0FJ ($t \leq 0$) (top left), 1b 0FJ ($t > 0$) (top right), 1b 1FJ ($t \leq 0$) (center left), 1b 1FJ ($t > 0$) (center right), 2b ($t \leq 0$) (bottom left), and 2b ($t > 0$) (bottom right), after a signal plus background fit across all channels, assuming a 100\GeV scalar mediator. The solid histograms for the simulated SM backgrounds are summed cumulatively and rescaled to luminosity, while the signal is shown as a dashed pink line. The grey dashed area in the upper panel represents the total uncertainty in the cumulative sum of the simulated SM backgrounds and signal, while in the lower panel it represents only the total uncertainty in the backgrounds. The data are represented by solid points with the horizontal bar indicating the width of the bin and the vertical one the associated statistical uncertainty. The last bin contains overflow events. The lower panel shows the ratio of data to post-fit background.}
\label{fig:sl:SRs:prepostfit}
\end{figure}

\begin{figure}[!htbp]
\centering
\includegraphics[width=0.49\textwidth]{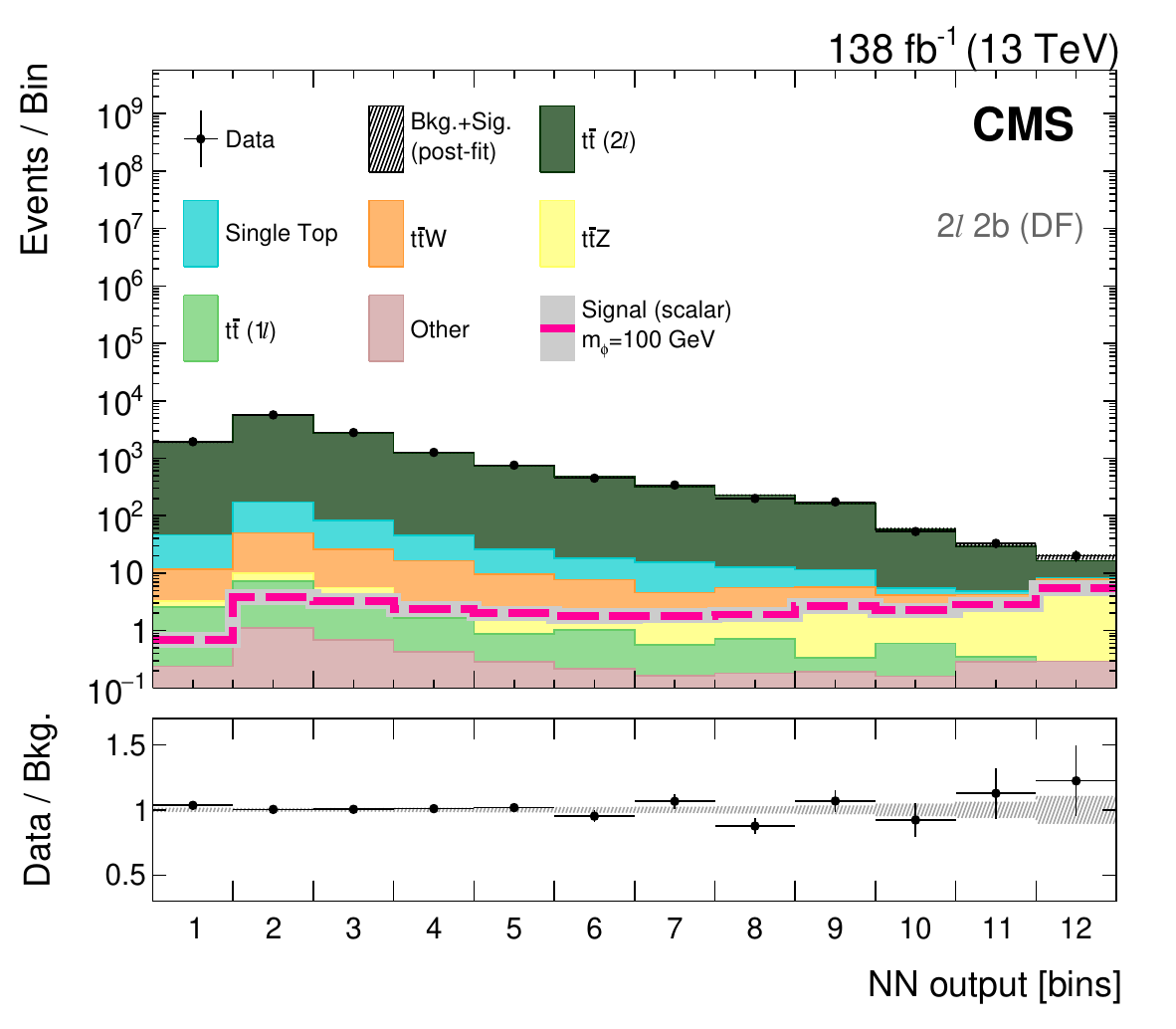}
\hfill
\includegraphics[width=0.49\textwidth]{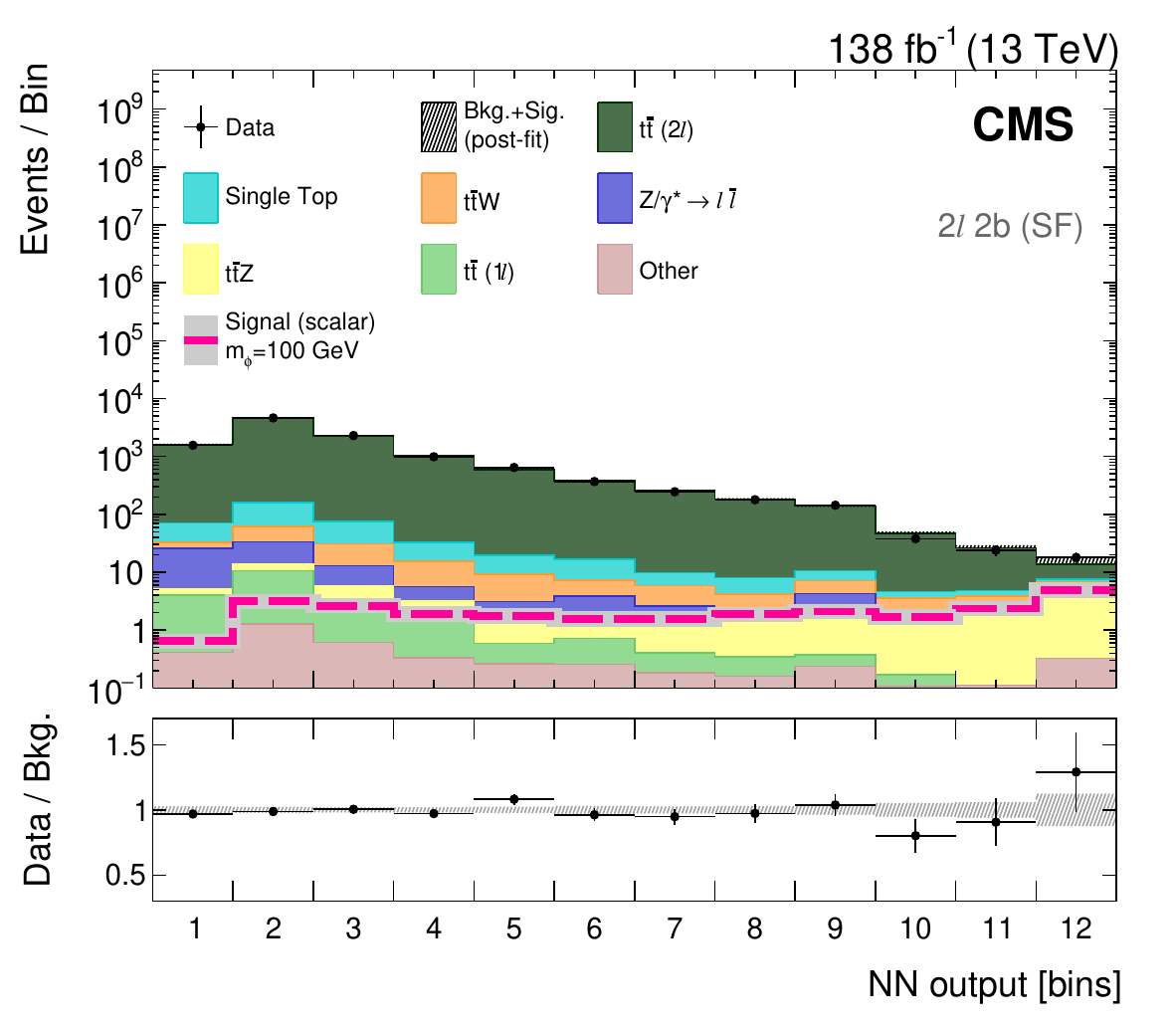}\\
\includegraphics[width=0.49\textwidth]{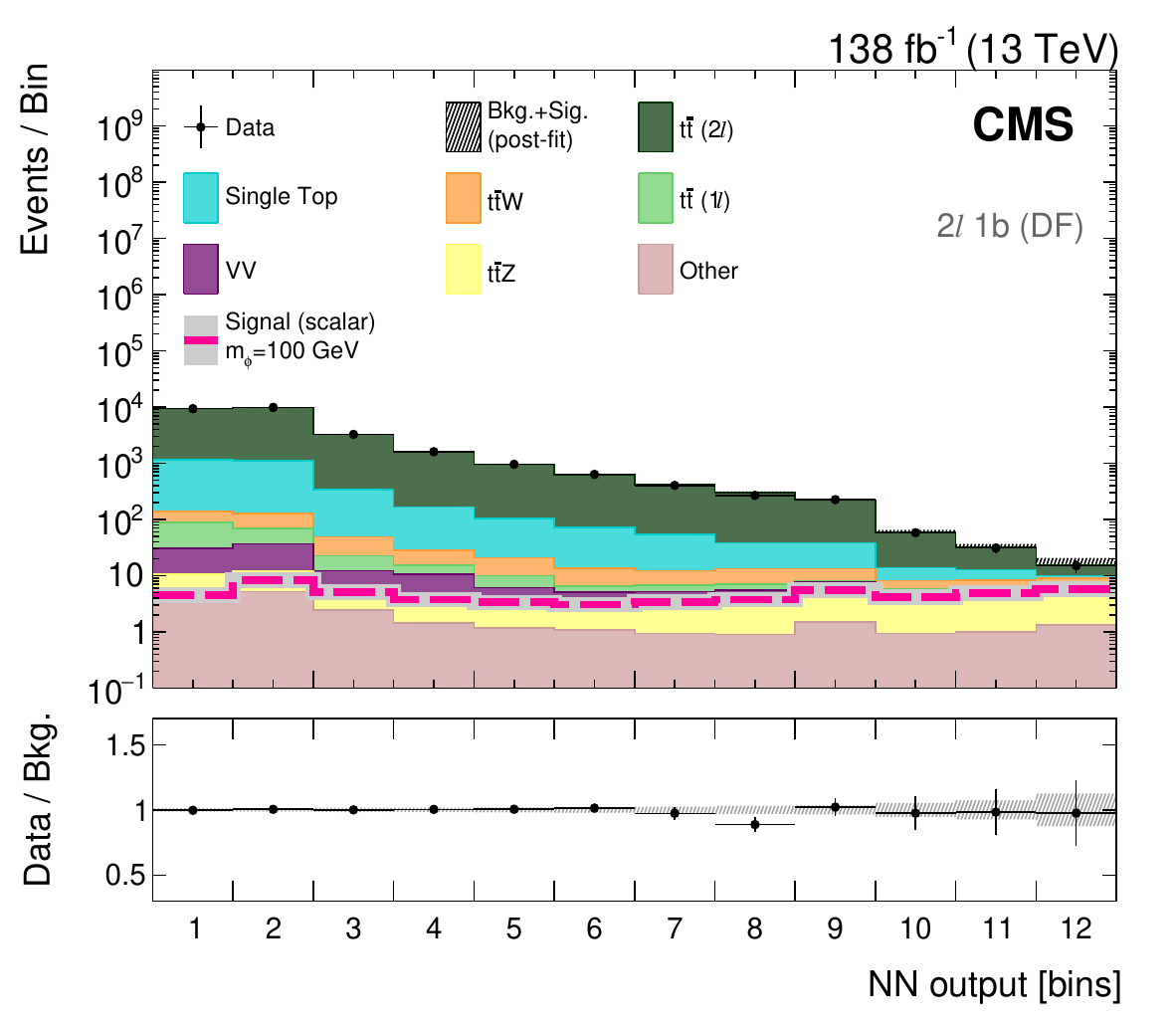}
\hfill
\includegraphics[width=0.49\textwidth]{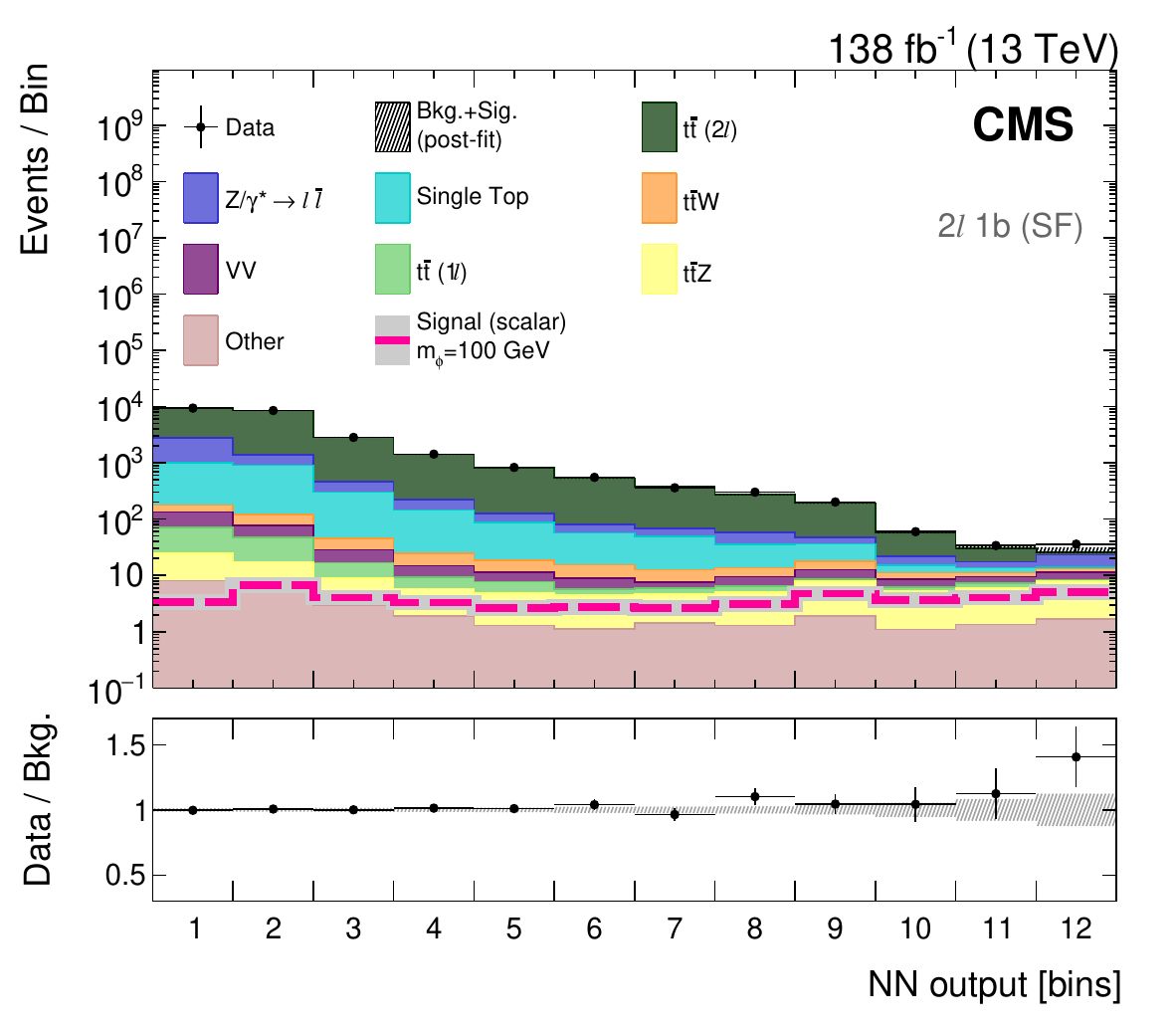}
\caption{The main discriminant distribution NN in the the four DL SRs: 2b (DF) (top left), 2b (SF) (top right), 1b (DF) (bottom left), and 1b (SF) (bottom right), after a signal plus background fit across all channels, assuming a 100\GeV scalar mediator. The solid histograms for the simulated SM backgrounds are summed cumulatively and rescaled to luminosity, while the signal is shown as a dashed pink line. The grey dashed area in the upper panel represents the total uncertainty in the cumulative sum of the simulated SM backgrounds and signal, while in the lower panel it represents only the total uncertainty in the backgrounds. The data are represented by solid points with the horizontal bar indicating the width of the bin and the vertical one the associated statistical uncertainty. The last bin contains overflow events. The lower panel shows the ratio of data to post-fit background.}
\label{fig:dl:SRs:prepostfit}
\end{figure}

A simultaneous fit to the binned \ptmiss and NN distributions is performed combining all of the aforementioned regions for each of the signal points considered. The results obtained for the fit to the 100\GeV scalar mediator (with best-fit signal strength $0.51\sigma_{\mathrm{th}}$) are illustrated in Figs. \ref{fig:ah:SRs:prepostfit}, \ref{fig:sl:SRs:prepostfit}, and \ref{fig:dl:SRs:prepostfit} for the AH, SL, and DL channels, respectively. The lower panels show the ratio of data and the SM prediction after performing the maximum likelihood fit. 

\section{Results}
\label{sec:results}
The results are interpreted using the signal model for scalar and pseudoscalar mediators with masses ranging from 50 to 500\GeV, with $\mchi=1\GeV$ and $g_{\PQq}=g_\chi=1$. The results are presented in terms of the signal strength parameter $\mu$, defined as the ratio between the measured and the theoretical cross sections for the \tDM and \ttDM production modes summed together. The theoretical cross sections for both signal models are obtained at LO.

The uncertainties from the AH and SL channels are found to be similar and dominate the combination, since these channels are most sensitive to the signal. In particular, the leading uncertainties are the b-tagging scale factor uncertainties (which can affect the number of \wjets and \Zinv events entering the SRs, and the uncertainties from the per-bin background estimation parameters, particularly for the \ttbar process. The next most significant uncertainties principally affect the DL channel: the normalization uncertainties of the \ttZ process and the last bin of the DY process, followed by the jet energy resolution uncertainties, which are observed to affect the shape of the \ttjets \mTiill distribution in particular.

For masses of the mediator particle below 200 (300)\GeV for the scalar (pseudoscalar) model, the leading contribution to the signal sensitivity stems from \ttDM processes because of their larger cross sections with respect to \tDM production. However, the \tDM cross section (\ptmiss distribution) drops less rapidly as a function of the mediator particle mass (\ptmiss) in comparison to the \ttDM.

\begin{figure}[htbp!]
\centering
\includegraphics[width=0.49\textwidth]{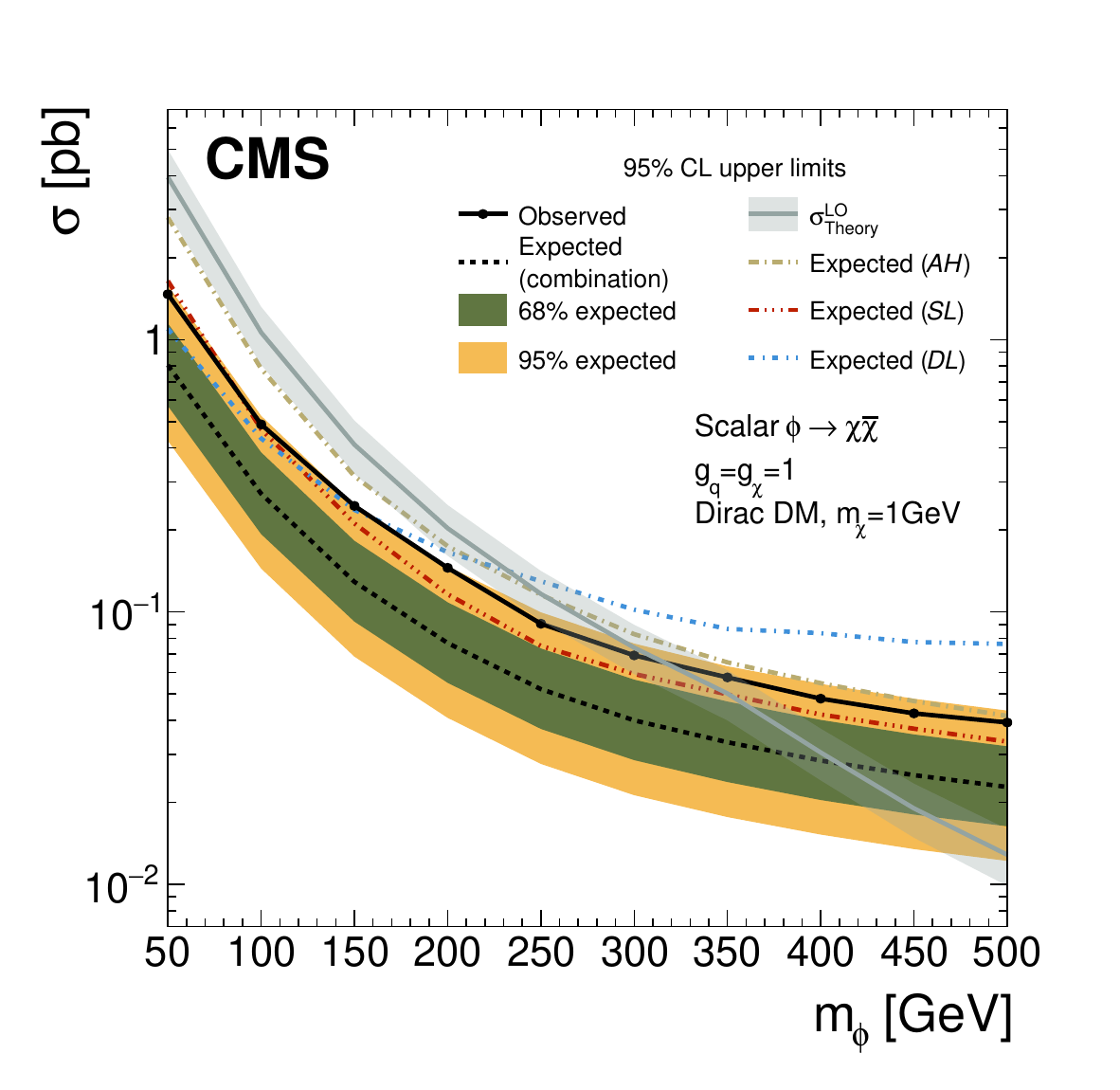}
\includegraphics[width=0.49\textwidth]{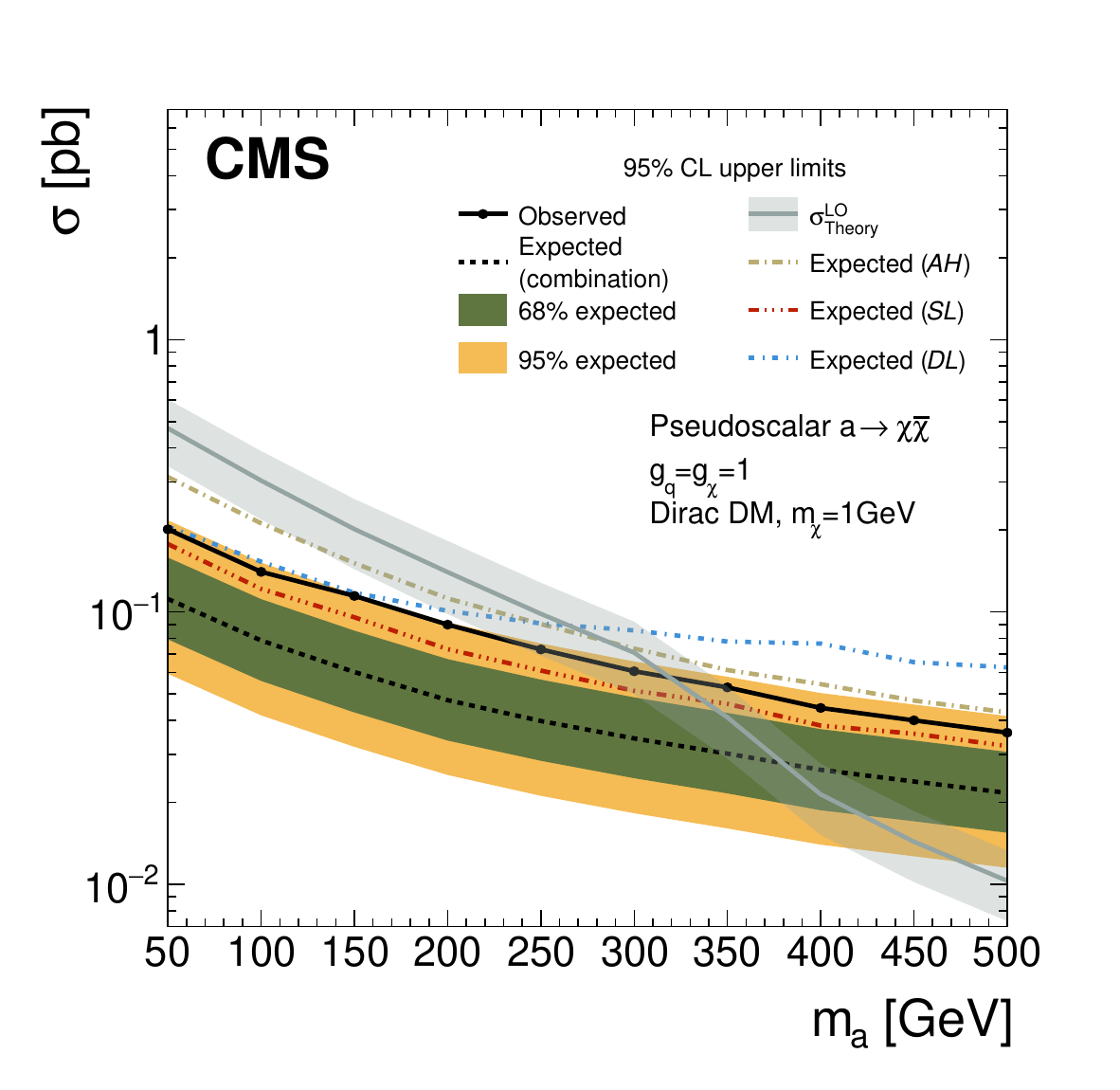}
\caption{
The model-independent 95\% \CL limits on production cross section for new physics processes for the scalar (left) and pseudoscalar (right) interactions. The expected limit is shown by the black dashed line with the 68 and 95\% \CL uncertainty bands in green and yellow, respectively, while the observed limit is shown by the solid black line. Theoretical LO cross section values for the DM model and their associated uncertainties are also presented (grey line).}
\label{fig:limits_modInd}
\end{figure}

Upper limits at 95\% confidence level (\CL) are computed using a modified frequentist approach with a test statistic based on the profile likelihood in the asymptotic approximation and the \CLs criterion~\cite{JUNK1999435,cls,Cowan2011}. The results are shown in Fig.~\ref{fig:limits_modInd} in terms of model-independent 95\% \CL limits on the production cross section for new physics processes for the DM scalar (left) and pseudoscalar (right) models. Cross section values ranging from 0.02 to 1\unit{pb} are probed depending on the assumption of the mediator of each new phenomenon. In terms of the DM model considered, we expect to exclude mediator masses below 410 (380)\GeV for the benchmark cross sections of the scalar (pseudoscalar) hypothesis.

A signal-like excess is observed in data. This excess is most visible in the AH 2b SR, and three of the DL SRs. The size of the excess is found to be statistically consistent across all SRs. Because the signal kinematic properties have limited correlations with the mass of the mediator, this excess is consistent with all mediator mass hypotheses, with local significances between 1.4 and 1.9 standard deviations. The largest significance is obtained for the 150\GeV pseudoscalar mediator. The variations in local significances are understood to come from the differing relative sensitivities of the lepton channels to the mediator mass hypotheses. In particular, the signal hypotheses with the lowest mediator masses (50 and 100\GeV) have slightly softer kinematic properties, which are less consistent with this excess. Because of the observed excess, we only exclude mediator masses below 310 (320)\GeV for the scalar (pseudoscalar) mediator

\begin{figure}[htbp!]
\centering
\includegraphics[width=0.49\textwidth]{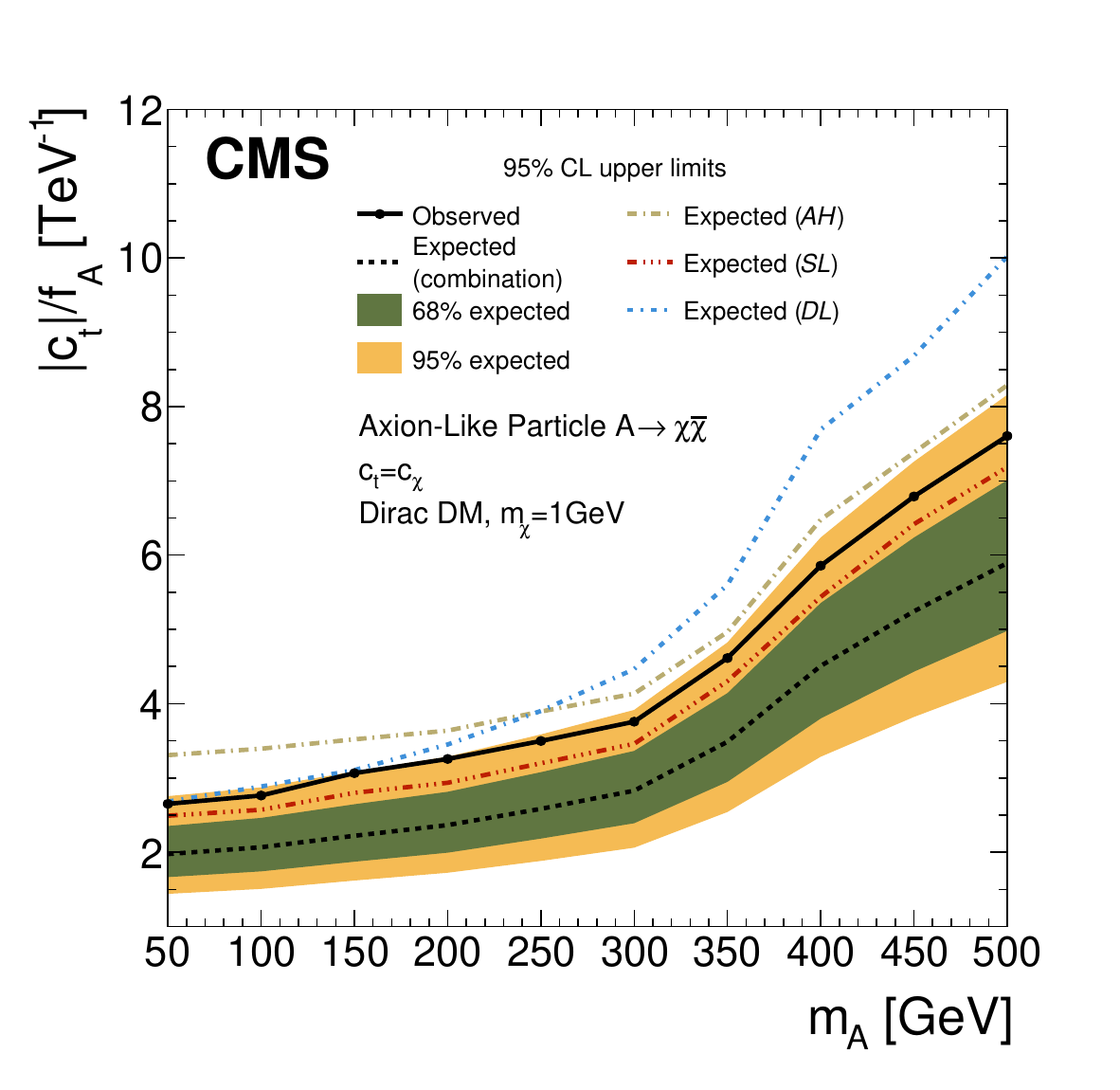}
\caption{
The 95\% \CL limits on the ratio of ALP-top coupling to the ALP decay constant ($\abs{c_{\mathrm{t}}}/f_{\mathrm{A}}$) as a function of mediator mass for the ALP mediator model. The expected limit is shown by the black dashed line with the 68 and 95\% \CL uncertainty bands in green and yellow, respectively, whereas the observed limit is shown by the solid black line.}
\label{fig:limits_ALP}
\end{figure}

As discussed in Section \ref{sec:introduction}, these signatures can also be interpreted in terms of models with an ALP mediator $\textrm{A}$. Since the additional pseudoscalar Higgs boson considered in Eq.~(\ref{eq:lagr_pseudoscalar}) and an ALP as introduced in Eq.~(\ref{eq:lagr_ALPs_pseudoscalar}) exhibit the same coupling structure, if the ALP has zero or negligible couplings to the gluon or electroweak gauge bosons~\cite{Anuar:2024myn}, the existing limits on the production cross section for pseudoscalar interactions as presented in Fig.~\ref{fig:limits_modInd} can be directly translated into upper limits on the ALP coupling to the top quark $c_\mathrm{t}$. Figure~\ref{fig:limits_ALP} shows the resulting limits on the ratio of the top quark coupling $c_{t}$ to the ALP scale $f_{A}$, as a function of the ALP mass $m_A$, assuming $c_t=c_\chi$. For ALP masses above 350\GeV, it is possible for the mediator to decay back to top quarks and hence varying $c_{t}$ will change the branching ratio of the ALP to DM; to resolve this ambiguity we take the coupling to DM $c_{\chi}$ equal to the coupling to top quarks $c_t$.

\section{Summary}
\label{sec:summary}
A search for dark matter (DM) produced in association with a single top quark or a top quark pair produced in interactions with a neutral scalar or pseudoscalar mediator in proton-proton collisions at a center-of-mass energy of 13\TeV has been presented. The search was performed using data corresponding to an integrated luminosity of 138\fbinv recorded by the CMS experiment between 2016 and 2018. This is the first search simultaneously optimized for the \tDM and \ttDM phase spaces across the zero lepton, single-lepton, and two lepton final states. In particular, lower jet multiplicities were considered with respect to the best LHC \ttDM searches, resulting in an increase in the sensitivity to processes where DM is produced in association with a single top quark.

The results are interpreted within a simplified model in which a scalar or pseudoscalar mediator couples to the top quark and subsequently decays into two DM particles. Scalar and pseudoscalar mediator masses below 410 and 380\GeV, respectively, are expected to be excluded at 95\% confidence level (\CL) assuming a DM particle mass of 1\GeV and mediator couplings to fermions and DM particles equal to unity. This work represents a 40\% improvement in sensitivity with respect to the previous combined CMS search for \tDM and \ttDM, which was the best LHC result with 2016 data. Of these improvements, up to 20\% comes from analysis enhancements and inclusion of the dileptonic final state, and 20\% from the larger data sample. A small signal-like excess is observed in data. Because the signal kinematic properties do not significantly depend on the mass of the mediator, this excess is consistent with all mediator mass hypotheses. The largest local significance for all mediator hypotheses is observed to be within two standard deviations. Because of this excess, mediator masses are only excluded below 310 (320)\GeV for the scalar (pseudoscalar) mediator. The results are also translated into model-independent 95\% \CL upper limits on the visible cross section of DM production in association with top quarks, ranging from 1\unit{pb} to 0.02\unit{pb}.

In addition, limits on the coupling of axion-like-particles (ALP) to top quarks are set for the first time. This is performed in the context of top quark(s) plus invisible signatures where the ALP couples to SM quarks as a mediator between the SM and fermionic DM particles.

\begin{acknowledgments}
We congratulate our colleagues in the CERN accelerator departments for the excellent performance of the LHC and thank the technical and administrative staffs at CERN and at other CMS institutes for their contributions to the success of the CMS effort. In addition, we gratefully acknowledge the computing centers and personnel of the Worldwide LHC Computing Grid and other centers for delivering so effectively the computing infrastructure essential to our analyses. Finally, we acknowledge the enduring support for the construction and operation of the LHC, the CMS detector, and the supporting computing infrastructure provided by the following funding agencies: SC (Armenia), BMBWF and FWF (Austria); FNRS and FWO (Belgium); CNPq, CAPES, FAPERJ, FAPERGS, and FAPESP (Brazil); MES and BNSF (Bulgaria); CERN; CAS, MoST, and NSFC (China); MINCIENCIAS (Colombia); MSES and CSF (Croatia); RIF (Cyprus); SENESCYT (Ecuador); ERC PRG, RVTT3 and MoER TK202 (Estonia); Academy of Finland, MEC, and HIP (Finland); CEA and CNRS/IN2P3 (France); SRNSF (Georgia); BMBF, DFG, and HGF (Germany); GSRI (Greece); NKFIH (Hungary); DAE and DST (India); IPM (Iran); SFI (Ireland); INFN (Italy); MSIP and NRF (Republic of Korea); MES (Latvia); LMTLT (Lithuania); MOE and UM (Malaysia); BUAP, CINVESTAV, CONACYT, LNS, SEP, and UASLP-FAI (Mexico); MOS (Montenegro); MBIE (New Zealand); PAEC (Pakistan); MES and NSC (Poland); FCT (Portugal); MESTD (Serbia); MCIN/AEI and PCTI (Spain); MOSTR (Sri Lanka); Swiss Funding Agencies (Switzerland); MST (Taipei); MHESI and NSTDA (Thailand); TUBITAK and TENMAK (Turkey); NASU (Ukraine); STFC (United Kingdom); DOE and NSF (USA).
       
\hyphenation{Rachada-pisek} Individuals have received support from the Marie-Curie program and the European Research Council and Horizon 2020 Grant, contract Nos.\ 675440, 724704, 752730, 758316, 765710, 824093, 101115353, 101002207, and COST Action CA16108 (European Union); the Leventis Foundation; the Alfred P.\ Sloan Foundation; the Alexander von Humboldt Foundation; the Science Committee, project no. 22rl-037 (Armenia); the Fonds pour la Formation \`a la Recherche dans l'Industrie et dans l'Agriculture (FRIA-Belgium); the Beijing Municipal Science \& Technology Commission, No. Z191100007219010 and Fundamental Research Funds for the Central Universities (China); the Ministry of Education, Youth and Sports (MEYS) of the Czech Republic; the Shota Rustaveli National Science Foundation, grant FR-22-985 (Georgia); the Deutsche Forschungsgemeinschaft (DFG), among others, under Germany's Excellence Strategy -- EXC 2121 ``Quantum Universe" -- 390833306, and under project number 400140256 - GRK2497; the Hellenic Foundation for Research and Innovation (HFRI), Project Number 2288 (Greece); the Hungarian Academy of Sciences, the New National Excellence Program - \'UNKP, the NKFIH research grants K 131991, K 133046, K 138136, K 143460, K 143477, K 146913, K 146914, K 147048, 2020-2.2.1-ED-2021-00181, TKP2021-NKTA-64, and 2021-4.1.2-NEMZ\_KI-2024-00036 (Hungary); the Council of Science and Industrial Research, India; ICSC -- National Research Center for High Performance Computing, Big Data and Quantum Computing and FAIR -- Future Artificial Intelligence Research, funded by the NextGenerationEU program (Italy); the Latvian Council of Science; the Ministry of Education and Science, project no. 2022/WK/14, and the National Science Center, contracts Opus 2021/41/B/ST2/01369 and 2021/43/B/ST2/01552 (Poland); the Funda\c{c}\~ao para a Ci\^encia e a Tecnologia, grant CEECIND/01334/2018 (Portugal); the National Priorities Research Program by Qatar National Research Fund; MCIN/AEI/10.13039/501100011033, ERDF ``a way of making Europe", and the Programa Estatal de Fomento de la Investigaci{\'o}n Cient{\'i}fica y T{\'e}cnica de Excelencia Mar\'{\i}a de Maeztu, grant MDM-2017-0765 and Programa Severo Ochoa del Principado de Asturias (Spain); the Chulalongkorn Academic into Its 2nd Century Project Advancement Project, and the National Science, Research and Innovation Fund via the Program Management Unit for Human Resources \& Institutional Development, Research and Innovation, grant B39G670016 (Thailand); the Kavli Foundation; the Nvidia Corporation; the SuperMicro Corporation; the Welch Foundation, contract C-1845; and the Weston Havens Foundation (USA).
\end{acknowledgments}

\bibliography{auto_generated}

\providecommand{\href}[2]{#2}\begingroup\raggedright\begin{thebibliography}{10}%
\makeatletter
\providecommand{\hrefCMSnoop }[0]{\@secondoftwo}%
\makeatother
\providecommand{\doi}{\texttt{doi:}\begingroup \urlstyle{tt}\Url}

\bibitem{Porter:2011nv}
\hrefCMSnoop {}{T.~A. Porter, R.~P. Johnson, and P.~W. Graham, ``Dark matter
  searches with astroparticle data'',} \textit{ Ann. Rev. Astron. Astrophys.}
  \textbf{ 49} (2011) 155,
  \href{http://dx.doi.org/10.1146/annurev-astro-081710-102528}{\doi{10.1146/annurev-astro-081710-102528}},
  \href{http://www.arXiv.org/abs/1104.2836}{\texttt{arXiv:1104.2836}}.

\bibitem{Bertone:2004pz}
\hrefCMSnoop {}{G.~Bertone, D.~Hooper, and J.~Silk, ``Particle dark matter:
  evidence, candidates and constraints'',} \textit{ Phys. Rept.} \textbf{ 405}
  (2005) 279,
  \href{http://dx.doi.org/10.1016/j.physrep.2004.08.031}{\doi{10.1016/j.physrep.2004.08.031}},
  \href{http://www.arXiv.org/abs/hep-ph/0404175}{\texttt{arXiv:hep-ph/0404175}}.

\bibitem{Abercrombie:2015wmb}
\hrefCMSnoop {}{D.~Abercrombie { et~al.}, ``Dark matter benchmark models for
  early {LHC} run-2 searches: Report of the {ATLAS}/{CMS} dark matter forum'',}
  \textit{ Phys. Dark Univ.} \textbf{ 27} (2020) 100371,
  \href{http://dx.doi.org/10.1016/j.dark.2019.100371}{\doi{10.1016/j.dark.2019.100371}},
  \href{http://www.arXiv.org/abs/1507.00966}{\texttt{arXiv:1507.00966}}.

\bibitem{PhysRevD.91.015017}
\hrefCMSnoop {}{M.~R. Buckley, D.~Feld, and D.~Goncalves, ``Scalar simplified
  models for dark matter'',} \textit{ Phys. Rev. D} \textbf{ 91} (2015)
  \href{http://dx.doi.org/10.1103/physrevd.91.015017}{\doi{10.1103/physrevd.91.015017}},
\href{http://www.arXiv.org/abs/1410.6497}{\texttt{arXiv:1410.6497}}.

\bibitem{PhysRevD.96.035031}
\hrefCMSnoop {}{D.~Pinna, A.~Zucchetta, M.~R. Buckley, and F.~Canelli, ``Single
  top quarks and dark matter'',} \textit{ Phys. Rev. D} \textbf{ 96} (2017)
  035031,
  \href{http://dx.doi.org/10.1103/PhysRevD.96.035031}{\doi{10.1103/PhysRevD.96.035031}},
\href{http://www.arXiv.org/abs/1701.05195}{\texttt{arXiv:1701.05195}}.

\bibitem{LHCDarkMatterWorkingGroup:2018ufk}
\hrefCMSnoop {}{{LHC Dark Matter Working Group} Collaboration, ``{LHC} dark
  matter working group: Next-generation spin-0 dark matter models'',} \textit{
  Phys. Dark Univ.} \textbf{ 27} (2020) 100351,
  \href{http://dx.doi.org/10.1016/j.dark.2019.100351}{\doi{10.1016/j.dark.2019.100351}},
  \href{http://www.arXiv.org/abs/1810.09420}{\texttt{arXiv:1810.09420}}.

\bibitem{Bauer:2017ota}
\hrefCMSnoop {}{M.~Bauer, U.~Haisch, and F.~Kahlhoefer, ``Simplified dark
  matter models with two {H}iggs doublets: I. pseudoscalar mediators'',}
  \textit{ JHEP} \textbf{ 05} (2017) 138,
  \href{http://dx.doi.org/10.1007/JHEP05(2017)138}{\doi{10.1007/JHEP05(2017)138}},
  \href{http://www.arXiv.org/abs/1701.07427}{\texttt{arXiv:1701.07427}}.

\bibitem{Peccei:1977hh}
\hrefCMSnoop {}{R.~D. Peccei and H.~R. Quinn, ``{CP} conservation in the
  presence of instantons'',} \textit{ Phys. Rev. Lett.} \textbf{ 38} (1977)
  1440,
  \href{http://dx.doi.org/10.1103/PhysRevLett.38.1440}{\doi{10.1103/PhysRevLett.38.1440}}.

\bibitem{Peccei:1977ur}
\hrefCMSnoop {}{R.~D. Peccei and H.~R. Quinn, ``Constraints imposed by {CP}
  conservation in the presence of instantons'',} \textit{ Phys. Rev. D}
  \textbf{ 16} (1977) 1791,
  \href{http://dx.doi.org/10.1103/PhysRevD.16.1791}{\doi{10.1103/PhysRevD.16.1791}}.

\bibitem{Weinberg:1977ma}
\hrefCMSnoop {}{S.~Weinberg, ``A new light boson?'',} \textit{ Phys. Rev.
  Lett.} \textbf{ 40} (1978) 223,
  \href{http://dx.doi.org/10.1103/PhysRevLett.40.223}{\doi{10.1103/PhysRevLett.40.223}}.

\bibitem{Wilczek:1977pj}
\hrefCMSnoop {}{F.~Wilczek, ``Problem of strong {$P$} and {$T$} invariance in
  the presence of instantons'',} \textit{ Phys. Rev. Lett.} \textbf{ 40} (1978)
  279,
  \href{http://dx.doi.org/10.1103/PhysRevLett.40.279}{\doi{10.1103/PhysRevLett.40.279}}.

\bibitem{Dolan:2014ska}
\hrefCMSnoop {}{M.~J. Dolan, F.~Kahlhoefer, C.~McCabe, and K.~Schmidt-Hoberg,
  ``A taste of dark matter: Flavour constraints on pseudoscalar mediators'',}
  \textit{ JHEP} \textbf{ 03} (2015) 171,
  \href{http://dx.doi.org/10.1007/JHEP03(2015)171}{\doi{10.1007/JHEP03(2015)171}},
  \href{http://www.arXiv.org/abs/1412.5174}{\texttt{arXiv:1412.5174}}.
  [Erratum: \DOI{10.1007/JHEP07(2015)103}].

\bibitem{Bharucha:2022lty}
\hrefCMSnoop {}{A.~Bharucha, F.~Br{\"u}mmer, N.~Desai, and S.~Mutzel,
  ``Axion-like particles as mediators for dark matter: beyond freeze-out'',}
  \textit{ JHEP} \textbf{ 02} (2023) 141,
  \href{http://dx.doi.org/10.1007/JHEP02(2023)141}{\doi{10.1007/JHEP02(2023)141}},
  \href{http://www.arXiv.org/abs/2209.03932}{\texttt{arXiv:2209.03932}}.

\bibitem{Allen:2024ndv}
S.~Allen\hrefCMSnoop {}{ { et~al.}, ``Electroweak axion portal to dark
  matter'',} 2024.
  \href{http://www.arXiv.org/abs/2405.02403}{\texttt{arXiv:2405.02403}}.

\bibitem{Blasi:2023hvb}
S.~Blasi\hrefCMSnoop {}{ { et~al.}, ``Top-philic {ALP} phenomenology at the
  {LHC}: the elusive mass-window'',} \textit{ JHEP} \textbf{ 06} (2024) 077,
  \href{http://dx.doi.org/10.1007/JHEP06(2024)077}{\doi{10.1007/JHEP06(2024)077}},
  \href{http://www.arXiv.org/abs/2311.16048}{\texttt{arXiv:2311.16048}}.

\bibitem{CMS:2021eha}
\hrefCMSnoop {}{{CMS Collaboration}, ``Combined searches for the production of
  supersymmetric top quark partners in proton\textendash{}proton collisions at
  $\sqrt{s} = 13$ {TeV}'',} \textit{ Eur. Phys. J. C} \textbf{ 81} (2021) 970,
  \href{http://dx.doi.org/10.1140/epjc/s10052-021-09721-5}{\doi{10.1140/epjc/s10052-021-09721-5}},
  \href{http://www.arXiv.org/abs/2107.10892}{\texttt{arXiv:2107.10892}}.

\bibitem{ATLAS-CONF-2022-007}
\hrefCMSnoop {}{{ATLAS Collaboration}, ``Constraints on spin-0 dark matter
  mediators and invisible {H}iggs decays using {ATLAS} 13 {TeV} pp collision
  data with two top quarks and missing transverse momentum in the final
  state'',} \textit{ Eur. Phys. J. C} \textbf{ 83} (2023) 503,
  \href{http://dx.doi.org/10.1140/epjc/s10052-023-11477-z}{\doi{10.1140/epjc/s10052-023-11477-z}},
  \href{http://www.arXiv.org/abs/2211.05426}{\texttt{arXiv:2211.05426}}.

\bibitem{ATLAS:2024rcx}
\hrefCMSnoop {}{{ATLAS Collaboration}, ``{Search for new phenomena with
  top-quark pairs and large missing transverse momentum using 140
  $\text{fb}^{-1}$ of pp collision data at $ \sqrt{s} $ = 13 TeV with the ATLAS
  detector}'',} \textit{ JHEP} \textbf{ 03} (2024) 139,
  \href{http://dx.doi.org/10.1007/JHEP03(2024)139}{\doi{10.1007/JHEP03(2024)139}},
  \href{http://www.arXiv.org/abs/2401.13430}{\texttt{arXiv:2401.13430}}.

\bibitem{ATLAS:2022znu}
\hrefCMSnoop {}{{ATLAS Collaboration}, ``{Search for dark matter produced in
  association with a single top quark and an energetic W boson in $\sqrt{s}=$
  13 TeV $pp$ collisions with the ATLAS detector}'',} \textit{ Eur. Phys. J. C}
  \textbf{ 83} (2023), no.~7, 603,
  \href{http://dx.doi.org/10.1140/epjc/s10052-023-11582-z}{\doi{10.1140/epjc/s10052-023-11582-z}},
  \href{http://www.arXiv.org/abs/2211.13138}{\texttt{arXiv:2211.13138}}.

\bibitem{ATLAS:2024kpy}
\hrefCMSnoop {}{{ATLAS Collaboration}, ``{Constraints on simplified dark matter
  models involving an $s$-channel mediator with the ATLAS detector in pp
  collisions at $\sqrt{s} = 13$~TeV}'',} \textit{ Eur. Phys. J. C} \textbf{ 84}
  (2024), no.~10, 1102,
  \href{http://dx.doi.org/10.1140/epjc/s10052-024-13215-5}{\doi{10.1140/epjc/s10052-024-13215-5}},
  \href{http://www.arXiv.org/abs/2404.15930}{\texttt{arXiv:2404.15930}}.

\bibitem{CMS-PAS-EXO-18-010}
\hrefCMSnoop {}{{CMS Collaboration}, ``Search for dark matter produced in
  association with a single top quark or a top quark pair in proton-proton
  collisions at $ \sqrt{s}=13 $ {TeV}'',} \textit{ JHEP} \textbf{ 03} (2019)
  141,
  \href{http://dx.doi.org/10.1007/JHEP03(2019)141}{\doi{10.1007/JHEP03(2019)141}},
  \href{http://www.arXiv.org/abs/1901.01553}{\texttt{arXiv:1901.01553}}.

\bibitem{hepdata}
\hrefCMSnoop {}{}{HEPD}ata record for this analysis, 2024.
\newblock
  \href{http://dx.doi.org/10.17182/hepdata.154755}{\doi{10.17182/hepdata.154755}}.

\bibitem{Chatrchyan:2008zzk}
\hrefCMSnoop {}{{CMS Collaboration}, ``The {CMS} experiment at the {CERN}
  {LHC}'',} \textit{ JINST} \textbf{ 3} (2008) S08004,
\href{http://dx.doi.org/10.1088/1748-0221/3/08/S08004}{\doi{10.1088/1748-0221/3/08/S08004}}.

\bibitem{Khachatryan:2016bia}
\hrefCMSnoop {}{{CMS Collaboration}, ``The {CMS} trigger system'',} \textit{
  JINST} \textbf{ 12} (2017) P01020,
  \href{http://dx.doi.org/10.1088/1748-0221/12/01/P01020}{\doi{10.1088/1748-0221/12/01/P01020}},
\href{http://www.arXiv.org/abs/1609.02366}{\texttt{arXiv:1609.02366}}.

\bibitem{CMS-PRF-14-001}
\hrefCMSnoop {}{{CMS Collaboration}, ``Particle-flow reconstruction and global
  event description with the {CMS} detector'',} \textit{ JINST} \textbf{ 12}
  (2017) P10003,
  \href{http://dx.doi.org/10.1088/1748-0221/12/10/P10003}{\doi{10.1088/1748-0221/12/10/P10003}},
\href{http://www.arXiv.org/abs/1706.04965}{\texttt{arXiv:1706.04965}}.

\bibitem{CMS-TDR-15-02}
\href {http://cds.cern.ch/record/2020886}{{CMS Collaboration}, ``Technical
  proposal for the {Phase-II} upgrade of the {Compact Muon Solenoid}'',} CMS
  Technical Proposal CERN-LHCC-2015-010, CMS-TDR-15-02, 2015.

\bibitem{Cacciari:2008gp}
\hrefCMSnoop {}{M.~Cacciari, G.~P. Salam, and G.~Soyez, ``The anti-\kt jet
  clustering algorithm'',} \textit{ JHEP} \textbf{ 04} (2008) 063,
  \href{http://dx.doi.org/10.1088/1126-6708/2008/04/063}{\doi{10.1088/1126-6708/2008/04/063}},
\href{http://www.arXiv.org/abs/0802.1189}{\texttt{arXiv:0802.1189}}.

\bibitem{Cacciari:2011ma}
\hrefCMSnoop {}{M.~Cacciari, G.~P. Salam, and G.~Soyez, ``{FastJet} user
  manual'',} \textit{ Eur. Phys. J. C} \textbf{ 72} (2012) 1896,
  \href{http://dx.doi.org/10.1140/epjc/s10052-012-1896-2}{\doi{10.1140/epjc/s10052-012-1896-2}},
\href{http://www.arXiv.org/abs/1111.6097}{\texttt{arXiv:1111.6097}}.

\bibitem{Cacciari:2008gn}
\hrefCMSnoop {}{M.~Cacciari, G.~P. Salam, and G.~Soyez, ``The catchment area of
  jets'',} \textit{ JHEP} \textbf{ 04} (2008) 005,
  \href{http://dx.doi.org/10.1088/1126-6708/2008/04/005}{\doi{10.1088/1126-6708/2008/04/005}},
\href{http://www.arXiv.org/abs/0802.1188}{\texttt{arXiv:0802.1188}}.

\bibitem{CMS-PAS-JME-16-003}
\href {https://cds.cern.ch/record/2256875}{{CMS Collaboration}, ``Jet
  algorithms performance in 13 {TeV} data'',} CMS Physics Analysis Summary
  CMS-PAS-JME-16-003, 2017.

\bibitem{CMS:2017wtu}
\hrefCMSnoop {}{{CMS Collaboration}, ``Identification of heavy-flavour jets
  with the {CMS} detector in pp collisions at 13 {TeV}'',} \textit{ JINST}
  \textbf{ 13} (2018) P05011,
  \href{http://dx.doi.org/10.1088/1748-0221/13/05/P05011}{\doi{10.1088/1748-0221/13/05/P05011}},
  \href{http://www.arXiv.org/abs/1712.07158}{\texttt{arXiv:1712.07158}}.

\bibitem{Nason:2004rx}
\hrefCMSnoop {}{P.~Nason, ``A new method for combining {NLO QCD} with shower
  {Monte Carlo} algorithms'',} \textit{ JHEP} \textbf{ 11} (2004) 040,
  \href{http://dx.doi.org/10.1088/1126-6708/2004/11/040}{\doi{10.1088/1126-6708/2004/11/040}},
\href{http://www.arXiv.org/abs/hep-ph/0409146}{\texttt{arXiv:hep-ph/0409146}}.

\bibitem{Frixione:2007vw}
\hrefCMSnoop {}{S.~Frixione, P.~Nason, and C.~Oleari, ``Matching {NLO QCD}
  computations with parton shower simulations: the {\POWHEG} method'',}
  \textit{ JHEP} \textbf{ 11} (2007) 070,
  \href{http://dx.doi.org/10.1088/1126-6708/2007/11/070}{\doi{10.1088/1126-6708/2007/11/070}},
\href{http://www.arXiv.org/abs/0709.2092}{\texttt{arXiv:0709.2092}}.

\bibitem{Alioli:2010xd}
\hrefCMSnoop {}{S.~Alioli, P.~Nason, C.~Oleari, and E.~Re, ``A general
  framework for implementing {NLO} calculations in shower {Monte Carlo}
  programs: the {\POWHEG box}'',} \textit{ JHEP} \textbf{ 06} (2010) 043,
  \href{http://dx.doi.org/10.1007/JHEP06(2010)043}{\doi{10.1007/JHEP06(2010)043}},
\href{http://www.arXiv.org/abs/1002.2581}{\texttt{arXiv:1002.2581}}.

\bibitem{Alioli:2009je}
\hrefCMSnoop {}{S.~Alioli, P.~Nason, C.~Oleari, and E.~Re, ``{NLO} single-top
  production matched with shower in {POWHEG}: s- and t-channel
  contributions'',} \textit{ JHEP} \textbf{ 09} (2009) 111,
  \href{http://dx.doi.org/10.1088/1126-6708/2009/09/111}{\doi{10.1088/1126-6708/2009/09/111}},
  \href{http://www.arXiv.org/abs/0907.4076}{\texttt{arXiv:0907.4076}}.
  [Erratum: \DOI{d10.1007/JHEP02(2010)011}].

\bibitem{Re:2010bp}
\hrefCMSnoop {}{E.~Re, ``Single-top {W}t-channel production matched with parton
  showers using the {POWHEG} method'',} \textit{ Eur. Phys. J. C} \textbf{ 71}
  (2011) 1547,
  \href{http://dx.doi.org/10.1140/epjc/s10052-011-1547-z}{\doi{10.1140/epjc/s10052-011-1547-z}},
  \href{http://www.arXiv.org/abs/1009.2450}{\texttt{arXiv:1009.2450}}.

\bibitem{Frixione:2007nw}
\hrefCMSnoop {}{S.~Frixione, P.~Nason, and G.~Ridolfi, ``A positive-weight
  next-to-leading-order {Monte Carlo} for heavy flavour hadroproduction'',}
  \textit{ JHEP} \textbf{ 09} (2007) 126,
  \href{http://dx.doi.org/10.1088/1126-6708/2007/09/126}{\doi{10.1088/1126-6708/2007/09/126}},
  \href{http://www.arXiv.org/abs/0707.3088}{\texttt{arXiv:0707.3088}}.

\bibitem{Czakon:2017wor}
M.~Czakon\hrefCMSnoop {}{ { et~al.}, ``Top-pair production at the {LHC} through
  {NNLO} {QCD} and {NLO} {EW}'',} \textit{ JHEP} \textbf{ 10} (2017) 186,
  \href{http://dx.doi.org/10.1007/JHEP10(2017)186}{\doi{10.1007/JHEP10(2017)186}},
  \href{http://www.arXiv.org/abs/1705.04105}{\texttt{arXiv:1705.04105}}.

\bibitem{Frederix:2012ps}
\hrefCMSnoop {}{R.~Frederix and S.~Frixione, ``Merging meets matching in
  {MC@NLO}'',} \textit{ JHEP} \textbf{ 12} (2012) 061,
  \href{http://dx.doi.org/10.1007/JHEP12(2012)061}{\doi{10.1007/JHEP12(2012)061}},
\href{http://www.arXiv.org/abs/1209.6215}{\texttt{arXiv:1209.6215}}.

\bibitem{Alwall:2014hca}
J.~Alwall\hrefCMSnoop {}{ { et~al.}, ``The automated computation of tree-level
  and next-to-leading order differential cross sections, and their matching to
  parton shower simulations'',} \textit{ JHEP} \textbf{ 07} (2014) 079,
  \href{http://dx.doi.org/10.1007/JHEP07(2014)079}{\doi{10.1007/JHEP07(2014)079}},
\href{http://www.arXiv.org/abs/1405.0301}{\texttt{arXiv:1405.0301}}.

\bibitem{Mangano:2006rw}
\hrefCMSnoop {}{M.~L. Mangano, M.~Moretti, F.~Piccinini, and M.~Treccani,
  ``Matching matrix elements and shower evolution for top-quark production in
  hadronic collisions'',} \textit{ JHEP} \textbf{ 01} (2007) 013,
  \href{http://dx.doi.org/10.1088/1126-6708/2007/01/013}{\doi{10.1088/1126-6708/2007/01/013}},
\href{http://www.arXiv.org/abs/hep-ph/0611129}{\texttt{arXiv:hep-ph/0611129}}.

\bibitem{Denner:2009gj}
\hrefCMSnoop {}{A.~Denner, S.~Dittmaier, T.~Kasprzik, and A.~Muck,
  ``Electroweak corrections to {W+jet} hadroproduction including leptonic
  {W-boson} decays'',} \textit{ JHEP} \textbf{ 08} (2009) 075,
  \href{http://dx.doi.org/10.1088/1126-6708/2009/08/075}{\doi{10.1088/1126-6708/2009/08/075}},
\href{http://www.arXiv.org/abs/0906.1656}{\texttt{arXiv:0906.1656}}.

\bibitem{Denner:2011vu}
\hrefCMSnoop {}{A.~Denner, S.~Dittmaier, T.~Kasprzik, and A.~Muck,
  ``Electroweak corrections to dilepton+jet production at hadron colliders'',}
  \textit{ JHEP} \textbf{ 06} (2011) 069,
  \href{http://dx.doi.org/10.1007/JHEP06(2011)069}{\doi{10.1007/JHEP06(2011)069}},
\href{http://www.arXiv.org/abs/1103.0914}{\texttt{arXiv:1103.0914}}.

\bibitem{Denner:2012ts}
\hrefCMSnoop {}{A.~Denner, S.~Dittmaier, T.~Kasprzik, and A.~Maeck,
  ``Electroweak corrections to monojet production at the {LHC}'',} \textit{
  Eur. Phys. J. C} \textbf{ 73} (2013) 2297,
  \href{http://dx.doi.org/10.1140/epjc/s10052-013-2297-x}{\doi{10.1140/epjc/s10052-013-2297-x}},
\href{http://www.arXiv.org/abs/1211.5078}{\texttt{arXiv:1211.5078}}.

\bibitem{Kuhn:2005gv}
\hrefCMSnoop {}{J.~H. Kuhn, A.~Kulesza, S.~Pozzorini, and M.~Schulze,
  ``Electroweak corrections to hadronic photon production at large transverse
  momenta'',} \textit{ JHEP} \textbf{ 03} (2006) 059,
  \href{http://dx.doi.org/10.1088/1126-6708/2006/03/059}{\doi{10.1088/1126-6708/2006/03/059}},
\href{http://www.arXiv.org/abs/hep-ph/0508253}{\texttt{arXiv:hep-ph/0508253}}.

\bibitem{Kallweit:2014xda}
S.~Kallweit\hrefCMSnoop {}{ { et~al.}, ``{NLO} electroweak automation and
  precise predictions for {W+multijet} production at the {LHC}'',} \textit{
  JHEP} \textbf{ 04} (2015) 012,
  \href{http://dx.doi.org/10.1007/JHEP04(2015)012}{\doi{10.1007/JHEP04(2015)012}},
\href{http://www.arXiv.org/abs/1412.5157}{\texttt{arXiv:1412.5157}}.

\bibitem{Kallweit:2015dum}
S.~Kallweit\hrefCMSnoop {}{ { et~al.}, ``{NLO QCD+EW} predictions for {V+jets}
  including off-shell vector-boson decays and multijet merging'',} \textit{
  JHEP} \textbf{ 04} (2016) 021,
  \href{http://dx.doi.org/10.1007/JHEP04(2016)021}{\doi{10.1007/JHEP04(2016)021}},
\href{http://www.arXiv.org/abs/1511.08692}{\texttt{arXiv:1511.08692}}.

\bibitem{Gehrmann:2014fva}
T.~Gehrmann\hrefCMSnoop {}{ { et~al.}, ``{$W^+W^-$} production at hadron
  colliders in next to next to leading order {QCD}'',} \textit{ Phys. Rev.
  Lett.} \textbf{ 113} (2014) 212001,
  \href{http://dx.doi.org/10.1103/PhysRevLett.113.212001}{\doi{10.1103/PhysRevLett.113.212001}},
  \href{http://www.arXiv.org/abs/1408.5243}{\texttt{arXiv:1408.5243}}.

\bibitem{Campbell:1999ah}
\hrefCMSnoop {}{J.~M. Campbell and R.~K. Ellis, ``An update on vector boson
  pair production at hadron colliders'',} \textit{ Phys. Rev. D} \textbf{ 60}
  (1999) 113006,
  \href{http://dx.doi.org/10.1103/PhysRevD.60.113006}{\doi{10.1103/PhysRevD.60.113006}},
  \href{http://www.arXiv.org/abs/hep-ph/9905386}{\texttt{arXiv:hep-ph/9905386}}.

\bibitem{Boveia:2016mrp}
\hrefCMSnoop {}{A.~Boveia { et~al.}, ``Recommendations on presenting {LHC}
  searches for missing transverse energy signals using simplified $s$-channel
  models of dark matter'',} \textit{ Phys. Dark Univ.} \textbf{ 27} (2020)
  100365,
  \href{http://dx.doi.org/10.1016/j.dark.2019.100365}{\doi{10.1016/j.dark.2019.100365}},
  \href{http://www.arXiv.org/abs/1603.04156}{\texttt{arXiv:1603.04156}}.

\bibitem{Ball:2014uwa}
\hrefCMSnoop {}{{NNPDF} Collaboration, ``Parton distributions for the {LHC Run
  II}'',} \textit{ JHEP} \textbf{ 04} (2015) 040,
  \href{http://dx.doi.org/10.1007/JHEP04(2015)040}{\doi{10.1007/JHEP04(2015)040}},
\href{http://www.arXiv.org/abs/1410.8849}{\texttt{arXiv:1410.8849}}.

\bibitem{NNPDF:2017mvq}
\hrefCMSnoop {}{{NNPDF} Collaboration, ``Parton distributions from
  high-precision collider data'',} \textit{ Eur. Phys. J. C} \textbf{ 77}
  (2017) 663,
  \href{http://dx.doi.org/10.1140/epjc/s10052-017-5199-5}{\doi{10.1140/epjc/s10052-017-5199-5}},
  \href{http://www.arXiv.org/abs/1706.00428}{\texttt{arXiv:1706.00428}}.

\bibitem{Sjostrand:2014zea}
T.~Sj{\"o}strand\hrefCMSnoop {}{ { et~al.}, ``An introduction to {PYTHIA}
  8.2'',} \textit{ Comput. Phys. Commun.} \textbf{ 191} (2015) 159,
  \href{http://dx.doi.org/10.1016/j.cpc.2015.01.024}{\doi{10.1016/j.cpc.2015.01.024}},
  \href{http://www.arXiv.org/abs/1410.3012}{\texttt{arXiv:1410.3012}}.

\bibitem{Khachatryan:2015pea}
\hrefCMSnoop {}{{CMS Collaboration}, ``Event generator tunes obtained from
  underlying event and multiparton scattering measurements'',} \textit{ Eur.
  Phys. J. C} \textbf{ 76} (2016) 155,
  \href{http://dx.doi.org/10.1140/epjc/s10052-016-3988-x}{\doi{10.1140/epjc/s10052-016-3988-x}},
  \href{http://www.arXiv.org/abs/1512.00815}{\texttt{arXiv:1512.00815}}.

\bibitem{CMS:2019csb}
\hrefCMSnoop {}{{CMS Collaboration}, ``Extraction and validation of a new set
  of {CMS} {PYTHIA8} tunes from underlying-event measurements'',} \textit{ Eur.
  Phys. J. C} \textbf{ 80} (2020) 4,
  \href{http://dx.doi.org/10.1140/epjc/s10052-019-7499-4}{\doi{10.1140/epjc/s10052-019-7499-4}},
  \href{http://www.arXiv.org/abs/1903.12179}{\texttt{arXiv:1903.12179}}.

\bibitem{Agostinelli:2002hh}
\hrefCMSnoop {}{{GEANT4} Collaboration, ``{\GEANTfour}---a simulation
  toolkit'',} \textit{ Nucl. Instrum. Meth. A} \textbf{ 506} (2003) 250,
\href{http://dx.doi.org/10.1016/S0168-9002(03)01368-8}{\doi{10.1016/S0168-9002(03)01368-8}}.

\bibitem{Khachatryan:2015hwa}
\hrefCMSnoop {}{{CMS Collaboration}, ``Performance of electron reconstruction
  and selection with the {CMS} detector in proton-proton collisions at
  $\sqrt{s} = 8$ {TeV}'',} \textit{ JINST} \textbf{ 10} (2015) P06005,
  \href{http://dx.doi.org/10.1088/1748-0221/10/06/P06005}{\doi{10.1088/1748-0221/10/06/P06005}},
\href{http://www.arXiv.org/abs/1502.02701}{\texttt{arXiv:1502.02701}}.

\bibitem{CMS-PAPER-MUO-10-004}
\hrefCMSnoop {}{{CMS Collaboration}, ``Performance of {CMS} muon reconstruction
  in pp collision events at $\sqrt{s} = 7$ {TeV}'',} \textit{ JINST} \textbf{
  7} (2012) P10002,
  \href{http://dx.doi.org/10.1088/1748-0221/7/10/P10002}{\doi{10.1088/1748-0221/7/10/P10002}},
\href{http://www.arXiv.org/abs/1206.4071}{\texttt{arXiv:1206.4071}}.

\bibitem{Bai:2012gs}
\hrefCMSnoop {}{Y.~Bai, H.-C. Cheng, J.~Gallicchio, and J.~Gu, ``Stop the top
  background of the stop search'',} \textit{ JHEP} \textbf{ 07} (2012) 110,
  \href{http://dx.doi.org/10.1007/JHEP07(2012)110}{\doi{10.1007/JHEP07(2012)110}},
\href{http://www.arXiv.org/abs/1203.4813}{\texttt{arXiv:1203.4813}}.

\bibitem{Graesser:2012modifiedtopness}
\hrefCMSnoop {}{{CMS Collaboration}, ``Search for direct top squark pair
  production in events with one lepton, jets, and missing transverse momentum
  at 13 {TeV} with the {CMS} experiment'',} \textit{ JHEP} \textbf{ 05} (2020)
  032,
  \href{http://dx.doi.org/10.1007/JHEP05(2020)032}{\doi{10.1007/JHEP05(2020)032}},
  \href{http://www.arXiv.org/abs/1912.08887}{\texttt{arXiv:1912.08887}}.

\bibitem{Lester:1999}
\hrefCMSnoop {}{C.~G. Lester and D.~J. Summers, ``Measuring masses of
  semiinvisibly decaying particles pair produced at hadron colliders'',}
  \textit{ Phys. Lett. B} \textbf{ 463} (1999) 99,
  \href{http://dx.doi.org/10.1016/S0370-2693(99)00945-4}{\doi{10.1016/S0370-2693(99)00945-4}},
  \href{http://www.arXiv.org/abs/hep-ph/9906349}{\texttt{arXiv:hep-ph/9906349}}.

\bibitem{Khachatryan:2015vra}
\hrefCMSnoop {}{{CMS Collaboration}, ``Searches for supersymmetry using the
  m$_{T2}$ variable in hadronic events produced in pp collisions at 8 {TeV}'',}
  \textit{ JHEP} \textbf{ 05} (2015) 078,
  \href{http://dx.doi.org/10.1007/JHEP05(2015)078}{\doi{10.1007/JHEP05(2015)078}},
  \href{http://www.arXiv.org/abs/1502.04358}{\texttt{arXiv:1502.04358}}.

\bibitem{Betchart:2013nba}
\hrefCMSnoop {}{B.~A. Betchart, R.~Demina, and A.~Harel, ``Analytic solutions
  for neutrino momenta in decay of top quarks'',} \textit{ Nucl. Instrum. Meth.
  A} \textbf{ 736} (2014) 169,
  \href{http://dx.doi.org/10.1016/j.nima.2013.10.039}{\doi{10.1016/j.nima.2013.10.039}},
  \href{http://www.arXiv.org/abs/1305.1878}{\texttt{arXiv:1305.1878}}.

\bibitem{CMS-LUM-17-003}
\hrefCMSnoop {}{{CMS Collaboration}, ``Precision luminosity measurement in
  proton-proton collisions at $\sqrt{s}={13\TeV}$ in 2015 and 2016 at {CMS}'',}
  \textit{ Eur. Phys. J. C} \textbf{ 81} (2021) 800,
  \href{http://dx.doi.org/10.1140/epjc/s10052-021-09538-2}{\doi{10.1140/epjc/s10052-021-09538-2}},
  \href{http://www.arXiv.org/abs/2104.01927}{\texttt{arXiv:2104.01927}}.

\bibitem{CMS-PAS-LUM-17-004}
\href {https://cds.cern.ch/record/2621960}{{CMS Collaboration}, ``{CMS}
  luminosity measurement for the 2017 data-taking period at $\sqrt{s} =
  13~\mathrm{TeV}$'',} technical report, CERN, Geneva, 2018.

\bibitem{CMS-PAS-LUM-18-002}
\href {https://cds.cern.ch/record/2676164}{{CMS Collaboration}, ``{CMS
  luminosity measurement for the 2018 data-taking period at $\sqrt{s} =
  13~\mathrm{TeV}$}'',} technical report, CERN, Geneva, 2019.

\bibitem{Butterworth:2015oua}
\hrefCMSnoop {}{J.~Butterworth { et~al.}, ``{PDF4LHC} recommendations for {LHC
  Run II}'',} \textit{ J. Phys. G} \textbf{ 43} (2016) 023001,
  \href{http://dx.doi.org/10.1088/0954-3899/43/2/023001}{\doi{10.1088/0954-3899/43/2/023001}},
\href{http://www.arXiv.org/abs/1510.03865}{\texttt{arXiv:1510.03865}}.

\bibitem{Sirunyan:2018nqx}
\hrefCMSnoop {}{{CMS Collaboration}, ``Measurement of the inelastic
  proton-proton cross section at $ \sqrt{s}=13 $ {TeV}'',} \textit{ JHEP}
  \textbf{ 07} (2018) 161,
  \href{http://dx.doi.org/10.1007/JHEP07(2018)161}{\doi{10.1007/JHEP07(2018)161}},
\href{http://www.arXiv.org/abs/1802.02613}{\texttt{arXiv:1802.02613}}.

\bibitem{Khachatryan:2016kdb}
\hrefCMSnoop {}{{CMS Collaboration}, ``Jet energy scale and resolution in the
  {CMS} experiment in pp collisions at 8 {TeV}'',} \textit{ JINST} \textbf{ 12}
  (2017) P02014,
  \href{http://dx.doi.org/10.1088/1748-0221/12/02/P02014}{\doi{10.1088/1748-0221/12/02/P02014}},
\href{http://www.arXiv.org/abs/1607.03663}{\texttt{arXiv:1607.03663}}.

\bibitem{CMS:2016sun}
\hrefCMSnoop {}{{CMS Collaboration}, ``Measurements of differential cross
  sections for associated production of a {W} boson and jets in proton-proton
  collisions at $\sqrt{s} = 8$ {TeV}'',} \textit{ Phys. Rev. D} \textbf{ 95}
  (2017) 052002,
  \href{http://dx.doi.org/10.1103/PhysRevD.95.052002}{\doi{10.1103/PhysRevD.95.052002}},
  \href{http://www.arXiv.org/abs/1610.04222}{\texttt{arXiv:1610.04222}}.

\bibitem{CMS:2016eha}
\hrefCMSnoop {}{{CMS Collaboration}, ``Measurement of the production cross
  section of a {W} boson in association with two b jets in pp collisions at
  $\sqrt{s} = 8$ {TeV}'',} \textit{ Eur. Phys. J. C} \textbf{ 77} (2017),
  no.~2, 92,
  \href{http://dx.doi.org/10.1140/epjc/s10052-016-4573-z}{\doi{10.1140/epjc/s10052-016-4573-z}},
  \href{http://www.arXiv.org/abs/1608.07561}{\texttt{arXiv:1608.07561}}.

\bibitem{CMS:2014bkk}
\hrefCMSnoop {}{{CMS Collaboration}, ``Measurements of jet multiplicity and
  differential production cross sections of {Z+jets} events in proton-proton
  collisions at $\sqrt{s} = 7$ {TeV}'',} \textit{ Phys. Rev. D} \textbf{ 91}
  (2015), no.~5, 052008,
  \href{http://dx.doi.org/10.1103/PhysRevD.91.052008}{\doi{10.1103/PhysRevD.91.052008}},
  \href{http://www.arXiv.org/abs/1408.3104}{\texttt{arXiv:1408.3104}}.

\bibitem{CMS:2016gmz}
\hrefCMSnoop {}{{CMS Collaboration}, ``Measurements of the associated
  production of a {Z} boson and b jets in pp collisions at $\sqrt{s} =
  8${TeV}'',} \textit{ Eur. Phys. J. C} \textbf{ 77} (2017), no.~11, 751,
  \href{http://dx.doi.org/10.1140/epjc/s10052-017-5140-y}{\doi{10.1140/epjc/s10052-017-5140-y}},
  \href{http://www.arXiv.org/abs/1611.06507}{\texttt{arXiv:1611.06507}}.

\bibitem{BARLOW1993219}
\hrefCMSnoop {}{R.~J. Barlow and C.~Beeston, ``Fitting using finite {Monte
  Carlo} samples'',} \textit{ Comput. Phys. Commun.} \textbf{ 77} (1993) 219,
\href{http://dx.doi.org/10.1016/0010-4655(93)90005-W}{\doi{10.1016/0010-4655(93)90005-W}}.

\bibitem{Khachatryan:2016mnb}
\hrefCMSnoop {}{{CMS Collaboration}, ``Measurement of differential cross
  sections for top quark pair production using the lepton+jets final state in
  proton-proton collisions at 13 {TeV}'',} \textit{ Phys. Rev. D} \textbf{ 95}
  (2017) 092001,
  \href{http://dx.doi.org/10.1103/PhysRevD.95.092001}{\doi{10.1103/PhysRevD.95.092001}},
\href{http://www.arXiv.org/abs/1610.04191}{\texttt{arXiv:1610.04191}}.

\bibitem{CAT-23-001}
\hrefCMSnoop {}{{CMS Collaboration}, ``The {CMS} statistical analysis and
  combination tool: {\textsc{Combine}}'',} \textit{ Comput. Softw. Big Sci.}
  \textbf{ 8} (2024) 19,
  \href{http://dx.doi.org/10.1007/s41781-024-00121-4}{\doi{10.1007/s41781-024-00121-4}},
  \href{http://www.arXiv.org/abs/2404.06614}{\texttt{arXiv:2404.06614}}.

\bibitem{Verkerke:2003ir}
\href
  {https://www.slac.stanford.edu/econf/C0303241/proc/papers/MOLT007.PDF}{W.~Verkerke
  and D.~Kirkby, ``The \textsc{RooFit} toolkit for data modeling'',} in
  \textit{ {Proc. 13th International Conference on Computing in High Energy and
  Nuclear Physics (CHEP 2003): La Jolla CA, United States, March 24--28,
  2003}}.
\newblock 2003.
\newblock
  \href{http://www.arXiv.org/abs/physics/0306116}{\texttt{arXiv:physics/0306116}}.
\newblock [eConf C0303241 (2003) MOLT007].

\bibitem{Moneta:2010pm}
L.~Moneta\hrefCMSnoop {}{ { et~al.}, ``The {R}oo{S}tats project'',} in \textit{
  13th International Workshop on Advanced Computing and Analysis Techniques in
  Physics Research (ACAT2010)}, T.~Speer { et~al.}, eds.
\newblock SISSA, 2010.
\newblock \href{http://www.arXiv.org/abs/1009.1003}{\texttt{arXiv:1009.1003}}.
\newblock {PoS(ACAT2010)057}.
  \href{http://dx.doi.org/10.22323/1.093.0057}{\doi{10.22323/1.093.0057}}.

\bibitem{CMS:2025448}
\hrefCMSnoop {}{``Dark sector searches with the {CMS} experiment'',} \textit{
  Physics Reports} \textbf{ 1115} (2025) 448,
  \href{http://dx.doi.org/10.1016/j.physrep.2024.09.013}{\doi{10.1016/j.physrep.2024.09.013}}.
  Included in the special issue {``ATLAS and CMS Run II Legacy results"}.

\bibitem{JUNK1999435}
\hrefCMSnoop {}{T.~Junk, ``Confidence level computation for combining searches
  with small statistics'',} \textit{ Nucl. Inst. Meth. A} \textbf{ 434} (1999)
  435,
  \href{http://dx.doi.org/10.1016/S0168-9002(99)00498-2}{\doi{10.1016/S0168-9002(99)00498-2}},
\href{http://www.arXiv.org/abs/hep-ex/9902226}{\texttt{arXiv:hep-ex/9902226}}.

\bibitem{cls}
\hrefCMSnoop {}{A.~L. Read, ``Presentation of search results: the {CL$_{s}$}
  technique'',} \textit{ J. Phys. G} \textbf{ 28} (2002) 2693,
\href{http://dx.doi.org/10.1088/0954-3899/28/10/313}{\doi{10.1088/0954-3899/28/10/313}}.

\bibitem{Cowan2011}
\hrefCMSnoop {}{G.~Cowan, K.~Cranmer, E.~Gross, and O.~Vitells, ``Asymptotic
  formulae for likelihood-based tests of new physics'',} \textit{ Eur. Phys. J.
  C} \textbf{ 71} (2011) 1554,
  \href{http://dx.doi.org/10.1140/epjc/s10052-011-1554-0}{\doi{10.1140/epjc/s10052-011-1554-0}},
  \href{http://www.arXiv.org/abs/1007.1727}{\texttt{arXiv:1007.1727}}.
[Erratum: \DOI{10.1140/epjc/s10052-013-2501-z}].

\bibitem{Anuar:2024myn}
A.~Anuar\hrefCMSnoop {}{ { et~al.}, ``{ALP}-ine quests at the {LHC}: hunting
  axion-like particles via peaks and dips in $t \bar{t}$ production'',} 2024.
  \href{http://www.arXiv.org/abs/2404.19014}{\texttt{arXiv:2404.19014}}.

\end{thebibliography}\endgroup
\cleardoublepage \appendix\section{The CMS Collaboration \label{app:collab}}\begin{sloppypar}\hyphenpenalty=5000\widowpenalty=500\clubpenalty=5000
\cmsinstitute{Yerevan Physics Institute, Yerevan, Armenia}
{\tolerance=6000
V.~Chekhovsky, A.~Hayrapetyan, V.~Makarenko\cmsorcid{0000-0002-8406-8605}, A.~Tumasyan\cmsAuthorMark{1}\cmsorcid{0009-0000-0684-6742}
\par}
\cmsinstitute{Institut f\"{u}r Hochenergiephysik, Vienna, Austria}
{\tolerance=6000
W.~Adam\cmsorcid{0000-0001-9099-4341}, J.W.~Andrejkovic, L.~Benato\cmsorcid{0000-0001-5135-7489}, T.~Bergauer\cmsorcid{0000-0002-5786-0293}, S.~Chatterjee\cmsorcid{0000-0003-2660-0349}, K.~Damanakis\cmsorcid{0000-0001-5389-2872}, M.~Dragicevic\cmsorcid{0000-0003-1967-6783}, P.S.~Hussain\cmsorcid{0000-0002-4825-5278}, M.~Jeitler\cmsAuthorMark{2}\cmsorcid{0000-0002-5141-9560}, N.~Krammer\cmsorcid{0000-0002-0548-0985}, A.~Li\cmsorcid{0000-0002-4547-116X}, D.~Liko\cmsorcid{0000-0002-3380-473X}, I.~Mikulec\cmsorcid{0000-0003-0385-2746}, J.~Schieck\cmsAuthorMark{2}\cmsorcid{0000-0002-1058-8093}, R.~Sch\"{o}fbeck\cmsAuthorMark{2}\cmsorcid{0000-0002-2332-8784}, D.~Schwarz\cmsorcid{0000-0002-3821-7331}, M.~Sonawane\cmsorcid{0000-0003-0510-7010}, W.~Waltenberger\cmsorcid{0000-0002-6215-7228}, C.-E.~Wulz\cmsAuthorMark{2}\cmsorcid{0000-0001-9226-5812}
% ...
\par}
\vskip\cmsinstskip
\dag:~Deceased\\
$^{1}$Also at Yerevan State University, Yerevan, Armenia\\
$^{2}$Also at TU Wien, Vienna, Austria\\
$^{3}$Also at Ghent University, Ghent, Belgium\\
$^{4}$Also at Universidade do Estado do Rio de Janeiro, Rio de Janeiro, Brazil\\
$^{5}$Also at FACAMP - Faculdades de Campinas, Sao Paulo, Brazil\\
$^{6}$Also at Universidade Estadual de Campinas, Campinas, Brazil\\
$^{7}$Also at Federal University of Rio Grande do Sul, Porto Alegre, Brazil\\
$^{8}$Also at University of Chinese Academy of Sciences, Beijing, China\\
$^{9}$Also at China Center of Advanced Science and Technology, Beijing, China\\
$^{10}$Also at University of Chinese Academy of Sciences, Beijing, China\\
$^{11}$Also at China Spallation Neutron Source, Guangdong, China\\
$^{12}$Now at Henan Normal University, Xinxiang, China\\
$^{13}$Also at University of Shanghai for Science and Technology, Shanghai, China\\
$^{14}$Now at The University of Iowa, Iowa City, Iowa, USA\\
$^{15}$Also at an institute formerly covered by a cooperation agreement with CERN\\
$^{16}$Also at Helwan University, Cairo, Egypt\\
$^{17}$Now at Zewail City of Science and Technology, Zewail, Egypt\\
$^{18}$Now at British University in Egypt, Cairo, Egypt\\
$^{19}$Now at Cairo University, Cairo, Egypt\\
$^{20}$Also at Purdue University, West Lafayette, Indiana, USA\\
$^{21}$Also at Universit\'{e} de Haute Alsace, Mulhouse, France\\
$^{22}$Also at Istinye University, Istanbul, Turkey\\
$^{23}$Also at an international laboratory covered by a cooperation agreement with CERN\\
$^{24}$Also at The University of the State of Amazonas, Manaus, Brazil\\
$^{25}$Also at University of Hamburg, Hamburg, Germany\\
$^{26}$Also at RWTH Aachen University, III. Physikalisches Institut A, Aachen, Germany\\
$^{27}$Also at Bergische University Wuppertal (BUW), Wuppertal, Germany\\
$^{28}$Also at Brandenburg University of Technology, Cottbus, Germany\\
$^{29}$Also at Forschungszentrum J\"{u}lich, Juelich, Germany\\
$^{30}$Also at CERN, European Organization for Nuclear Research, Geneva, Switzerland\\
$^{31}$Also at HUN-REN ATOMKI - Institute of Nuclear Research, Debrecen, Hungary\\
$^{32}$Now at Universitatea Babes-Bolyai - Facultatea de Fizica, Cluj-Napoca, Romania\\
$^{33}$Also at MTA-ELTE Lend\"{u}let CMS Particle and Nuclear Physics Group, E\"{o}tv\"{o}s Lor\'{a}nd University, Budapest, Hungary\\
$^{34}$Also at HUN-REN Wigner Research Centre for Physics, Budapest, Hungary\\
$^{35}$Also at Physics Department, Faculty of Science, Assiut University, Assiut, Egypt\\
$^{36}$Also at Punjab Agricultural University, Ludhiana, India\\
$^{37}$Also at University of Visva-Bharati, Santiniketan, India\\
$^{38}$Also at Indian Institute of Science (IISc), Bangalore, India\\
$^{39}$Also at Amity University Uttar Pradesh, Noida, India\\
$^{40}$Also at IIT Bhubaneswar, Bhubaneswar, India\\
$^{41}$Also at Institute of Physics, Bhubaneswar, India\\
$^{42}$Also at University of Hyderabad, Hyderabad, India\\
$^{43}$Also at Deutsches Elektronen-Synchrotron, Hamburg, Germany\\
$^{44}$Also at Isfahan University of Technology, Isfahan, Iran\\
$^{45}$Also at Sharif University of Technology, Tehran, Iran\\
$^{46}$Also at Department of Physics, University of Science and Technology of Mazandaran, Behshahr, Iran\\
$^{47}$Also at Department of Physics, Faculty of Science, Arak University, ARAK, Iran\\
$^{48}$Also at Italian National Agency for New Technologies, Energy and Sustainable Economic Development, Bologna, Italy\\
$^{49}$Also at Centro Siciliano di Fisica Nucleare e di Struttura Della Materia, Catania, Italy\\
$^{50}$Also at Universit\`{a} degli Studi Guglielmo Marconi, Roma, Italy\\
$^{51}$Also at Scuola Superiore Meridionale, Universit\`{a} di Napoli 'Federico II', Napoli, Italy\\
$^{52}$Also at Fermi National Accelerator Laboratory, Batavia, Illinois, USA\\
$^{53}$Also at Lulea University of Technology, Lulea, Sweden\\
$^{54}$Also at Consiglio Nazionale delle Ricerche - Istituto Officina dei Materiali, Perugia, Italy\\
$^{55}$Also at Institut de Physique des 2 Infinis de Lyon (IP2I ), Villeurbanne, France\\
$^{56}$Also at Department of Applied Physics, Faculty of Science and Technology, Universiti Kebangsaan Malaysia, Bangi, Malaysia\\
$^{57}$Also at Consejo Nacional de Ciencia y Tecnolog\'{i}a, Mexico City, Mexico\\
$^{58}$Also at Trincomalee Campus, Eastern University, Sri Lanka, Nilaveli, Sri Lanka\\
$^{59}$Also at Saegis Campus, Nugegoda, Sri Lanka\\
$^{60}$Also at National and Kapodistrian University of Athens, Athens, Greece\\
$^{61}$Also at Ecole Polytechnique F\'{e}d\'{e}rale Lausanne, Lausanne, Switzerland\\
$^{62}$Also at Universit\"{a}t Z\"{u}rich, Zurich, Switzerland\\
$^{63}$Also at Stefan Meyer Institute for Subatomic Physics, Vienna, Austria\\
$^{64}$Also at Laboratoire d'Annecy-le-Vieux de Physique des Particules, IN2P3-CNRS, Annecy-le-Vieux, France\\
$^{65}$Also at Near East University, Research Center of Experimental Health Science, Mersin, Turkey\\
$^{66}$Also at Konya Technical University, Konya, Turkey\\
$^{67}$Also at Izmir Bakircay University, Izmir, Turkey\\
$^{68}$Also at Adiyaman University, Adiyaman, Turkey\\
$^{69}$Also at Bozok Universitetesi Rekt\"{o}rl\"{u}g\"{u}, Yozgat, Turkey\\
$^{70}$Also at Marmara University, Istanbul, Turkey\\
$^{71}$Also at Milli Savunma University, Istanbul, Turkey\\
$^{72}$Also at Kafkas University, Kars, Turkey\\
$^{73}$Now at Istanbul Okan University, Istanbul, Turkey\\
$^{74}$Also at Hacettepe University, Ankara, Turkey\\
$^{75}$Also at Erzincan Binali Yildirim University, Erzincan, Turkey\\
$^{76}$Also at Istanbul University -  Cerrahpasa, Faculty of Engineering, Istanbul, Turkey\\
$^{77}$Also at Yildiz Technical University, Istanbul, Turkey\\
$^{78}$Also at School of Physics and Astronomy, University of Southampton, Southampton, United Kingdom\\
$^{79}$Also at IPPP Durham University, Durham, United Kingdom\\
$^{80}$Also at Monash University, Faculty of Science, Clayton, Australia\\
$^{81}$Also at Universit\`{a} di Torino, Torino, Italy\\
$^{82}$Also at Bethel University, St. Paul, Minnesota, USA\\
$^{83}$Also at Karamano\u {g}lu Mehmetbey University, Karaman, Turkey\\
$^{84}$Also at California Institute of Technology, Pasadena, California, USA\\
$^{85}$Also at United States Naval Academy, Annapolis, Maryland, USA\\
$^{86}$Also at Ain Shams University, Cairo, Egypt\\
$^{87}$Also at Bingol University, Bingol, Turkey\\
$^{88}$Also at Georgian Technical University, Tbilisi, Georgia\\
$^{89}$Also at Sinop University, Sinop, Turkey\\
$^{90}$Also at Erciyes University, Kayseri, Turkey\\
$^{91}$Also at Horia Hulubei National Institute of Physics and Nuclear Engineering (IFIN-HH), Bucharest, Romania\\
$^{92}$Now at another institute formerly covered by a cooperation agreement with CERN\\
$^{93}$Also at Texas A\&M University at Qatar, Doha, Qatar\\
$^{94}$Also at Kyungpook National University, Daegu, Korea\\
$^{95}$Also at Yerevan Physics Institute, Yerevan, Armenia\\
$^{96}$Also at another international laboratory covered by a cooperation agreement with CERN\\
$^{97}$Also at Imperial College, London, United Kingdom\\
$^{98}$Also at Institute of Nuclear Physics of the Uzbekistan Academy of Sciences, Tashkent, Uzbekistan\\
$^{99}$Also at another institute formerly covered by a cooperation agreement with CERN\\
\end{sloppypar}
\end{document}